\documentclass[aps,prb,superscriptaddress,a4paper,amsmath,amssymb,preprint,floatfix]{revtex4-2}

\usepackage{amsmath}
\usepackage{amsfonts}
\usepackage{amssymb}
\usepackage{natbib}
\usepackage{graphicx}
\usepackage{array}

\begin{document}

\title{Power Transfer in Magnetoelectric Resonators: a Combined Analytical and Finite Element Study}
\author{Emma Van Meirvenne}
\affiliation{Imec, 3001 Leuven, Belgium}
\affiliation{KU Leuven, Departement Elektrotechniek, 3001 Leuven, Belgium}

\author{Frederic Vanderveken}
\affiliation{Imec, 3001 Leuven, Belgium}
\affiliation{KU Leuven, Departement Elektrotechniek, 3001 Leuven, Belgium}

\author{Daniele Narducci}
\affiliation{Imec, 3001 Leuven, Belgium}
\affiliation{KU Leuven, Departement Elektrotechniek, 3001 Leuven, Belgium}

\author{Bart Sor\'ee} 
\affiliation{Imec, 3001 Leuven, Belgium}
\affiliation{KU Leuven, Departement Elektrotechniek, 3001 Leuven, Belgium}

\author{Florin Ciubotaru}
\affiliation{Imec, 3001 Leuven, Belgium}

\author{Christoph Adelmann}
\affiliation{Imec, 3001 Leuven, Belgium}

\begin{abstract}
	We present an analytical model for power transfer in a magnetoelectric film bulk acoustic resonator (FBAR) comprising a piezoelectric-magnetostrictive bilayer. The model describes the power flow between the elastic and magnetic systems, quantifying the transduction efficiency when the FBAR operates as a magnetic transducer. By applying the model to example systems using piezoelectric ScAlN and magnetostrictive CoFeB, Ni, or Terfenol-D layers, we demonstrate the potential for achieving high efficiencies in magnetoelectric transducers, rendering them ideal for efficient ferromagnetic resonance excitation. The validity of the model's assumptions is confirmed through comparison with a numerical finite element resonator model in COMSOL\texttrademark. The finite element model further enables a comprehensive study of the resonator's dynamic behavior, including transient and steady-state regimes, and the identification of resonant frequencies within the system.
\end{abstract}

\maketitle

\clearpage

\section{Introduction}

\noindent In recent years, magnetoelectric transducers, capable of efficiently converting electrical signals into magnetic excitations and \textit{vice versa}, have garnered significant attention both from an applied as well as a fundamental point of view. Particularly within the emerging field of magnonics \cite{1_ChumakA.V.2015Ms,2_MahmoudAbdulqader2020Itsw,3_BarmanAnjan2021T2MR,4_ChumakA.V.2022AiMR}, which uses spin waves as information carriers, these transducers are considered essential for enabling ultralow-power data processing \cite{2_MahmoudAbdulqader2020Itsw,5_KhitunAlexander2011Nmlc,6_DienyB.2020Oacf,incorvia_spintronics_2024}. While various spin-wave transducers have been proposed and realized, including inductive antennas \cite{7_citation-key,8_ChumakA.V.2009Spia,9_CiubotaruF.2016Aeps,10_ConnellyDavidA.2021Eetf,11_VandervekenFrederic2022Lcmf}, as well as devices based on spin-orbit torques \cite{12_DivinskiyBoris2018EaAo,13_TalmelliGiacomo2018SEbS}, or spin-transfer torques \cite{14_MadamiM.2011Dooa}, they primarily rely on current-based mechanisms. This inevitably leads to Joule heating in the transducers, strongly limiting their energy efficiency, especially when miniaturized \cite{11_VandervekenFrederic2022Lcmf}.

In contrast, magnetoelectric transducers leverage the coupling of piezoelectric and magnetostrictive effects in composite materials. These devices are driven by voltage signals, and therefore promise significantly higher energy efficiency at the nanoscale compared to current-based transducers \cite{2_MahmoudAbdulqader2020Itsw,5_KhitunAlexander2011Nmlc,15_KhitunAlexander2009Mswa,incorvia_spintronics_2024}. In this configuration, AC voltage signals induce elastodynamic excitations via the piezoelectric effect, which subsequently couple to the magnetization through magnetostriction. Recent research has extensively explored the dynamic interplay between magnetization and elastodynamic modes \cite{16_BhuktareSwapnil2017GBoM,17_VerbaRoman2021PNoM,18_BabuNandanK.P2021TIbS,19_PolzikovaN.I.2016Aspi,20_AlekseevS.G.2020Mppi,21_BichurinM.I.2010Psot,22_LitvinenkoA.2021TMOw}.

Several approaches have been pursued to induce magnetization dynamics through dynamic strain. Recent research has primarily focused on coupling magnetization dynamics with surface acoustic waves (SAWs) generated in piezoelectric substrates by interdigitated transducers \cite{16_BhuktareSwapnil2017GBoM,17_VerbaRoman2021PNoM,23_CherepovSergiy2014Eswg,24_FoersterMichael2017Diod,25_KittmannAnne2018WBLN,26_CastillaDavid2020Mpoa}. While these structures offer potential for efficient magnetoelastic coupling \cite{27_MazzamurroAurelien2020GMCi}, achieving the multi-GHz frequencies required for strong coupling remains challenging. Additionally, scaling SAW devices to the nanoscale, essential for spintronic applications \cite{2_MahmoudAbdulqader2020Itsw,6_DienyB.2020Oacf}, is difficult. In contrast, devices based on bulk acoustic waves, which can generate elasto- and magnetodynamic excitations through voltage application to a piezoelectric-magnetostrictive bilayer \cite{28_DuttaSourav2014SCMo,29_PertsevNA2008GMEv,30_BalinskiyMichael2018MSWM,31_VandervekenFrederic2020Eapo}, have received less attention despite their potential for higher-frequency operation and better scalability.

Magnetoelectric coupling can be significantly enhanced through the exploitation of mechanical resonances within the system. High-overtone bulk acoustic resonators (HBARs) incorporating an additional magnetostrictive layer have been extensively modeled and studied experimentally, demonstrating large magnetoelectric coupling at HBAR resonances \cite{19_PolzikovaN.I.2016Aspi,20_AlekseevS.G.2020Mppi,21_BichurinM.I.2010Psot,22_LitvinenkoA.2021TMOw,32_PolzikovaN.I.2018EEoS,33_PolzikovaN.I.2018Famf,34_PolzikovaN.I.2019Aeae,35_PolzikovaN.I.2019RSPi,36_AlekseevS.G.2019YIGT}. However, from a practical standpoint, film bulk acoustic resonators (FBARs), for example as solidly mounted resonators (SMRs) \cite{37_LakinK.M.1995Smra,38_WeinsteinDana2010TRBT}, are more suitable for spintronic applications. Consequently, FBARs have been proposed as spin-wave transducers for magnonic logic \cite{39_KhitunAlexander2007Fsol,5_KhitunAlexander2011Nmlc}, but a comprehensive understanding of their efficiency, scaling characteristics, and dependence on material parameters remains elusive. While previous studies have explored magnetoelastic coupling and resonance effects in magnetoelectric FBARs \cite{21_BichurinM.I.2010Psot,41_BichurinM.I.2012MoMI}, a detailed analysis of power flow, device scalability, and transducer efficiency is still lacking.

In this work, we present a theoretical framework describing magnetoelastic coupling and power flow within a magnetoelectric (piezoelectric--magnetostrictive) FBAR at concurrent mechanical and magnetic resonances. Our analytical model enables the derivation of expressions for elastic and magnetic power losses, as well as the power transfer from the elastic into the magnetic domain, facilitating the assessment of the transducer efficiency. A complementary finite element method (FEM) model is developed to validate assumptions inherent to the analytical model \cite{1Comsol_mm,2comsol,3comsol}. Finally, a case study, employing example systems, is used to assess the impact of material and dimensional parameters on transducer performance, demonstrating the potential for achieving high transduction efficiencies in magnetoelectric FBARs.

\section{Analytical magnetoelectric FBAR Model} \label{sec:boundary_value_problem}

\subsection{Resonator geometry and configuration}

\begin{figure}[p] 
	\centering
	\includegraphics[width=8cm]{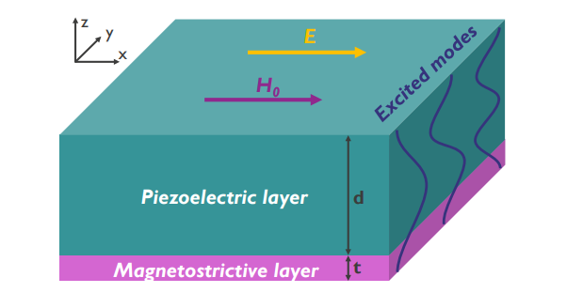}
	\caption{Schematic illustration of the magnetoelectric resonator structure along with the associated coordinate system. The applied time-varying electric field $E$ excites elastic waves through the piezoelectric layer, while a static external magnetic field $H_0$ is applied to enable resonant magnetoelastic coupling to FMR. The diagram also depicts different shear elastic wave modes along the thickness of the resonator.}
	\label{Layout_MEResonatorr}
\end{figure}

\noindent The studied structure comprises a piezoelectric-magnetostrictive bilayer with a total thickness of $d+t$, as illustrated in Fig.~\ref{Layout_MEResonatorr}. Assuming that the lateral dimensions are much larger than the total thickness, the system can be approximated as one-dimensional, with spatial variations in magnetization and displacement only in the $z$-direction. We further impose free elastic boundary conditions at the top and bottom surfaces of the resonator, analogous to an ideal FBAR device \cite{42_alma999647780101488}. The resonator is actuated by applying a AC electric field across the piezoelectric element, inducing dynamic strain within the bilayer. This strain then propagates into the magnetostrictive layer and excites magnetization dynamics.

Several configurations exist for the orientation of magnetization and the application of electric fields, leading to varying degrees of coupling between elastic and magnetic domains. While some configurations exhibit weak second-order coupling only, others offer strong first-order coupling. In this work, we investigate a configuration where the electric field is applied laterally along the $x$-direction, parallel to the static magnetization, as depicted in Fig.~\ref{Layout_MEResonatorr}. This configuration enables the excitation of confined elastic shear wave resonator modes along the thickness of the structure, resulting in efficient first-order magnetoelastic coupling \cite{31_VandervekenFrederic2020Eapo,43_DuflouRutger2017Msom}.

\subsection{Boundary value problem} 

\noindent Next, we derive the differential equations governing the dynamics within the piezoelectric and magnetostrictive media. Subsequently, we specify the boundary conditions and present solutions to the resulting boundary value problem. These solutions will serve as the foundation for calculating magnetic and elastic power absorption, as well as transducer efficiency.

\subsubsection{Equations of motion in the piezoelectric film} 
\noindent The constitutive relations describing the electromechanical coupling in the piezoelectric medium are given by \cite{42_alma999647780101488,44_alma9992196915901488}
\begin{align}
    \label{eq_piezoelectric}
    \sigma_{ij} & = \sum_{k,l} c_{ijkl}^{E}S_{kl} - \sum_{k}e_{kij}E_{k}, \\ 
    D_{i} & = \sum_{k,l} e_{ikl}S_{kl} + \sum_{j}\varepsilon_{ij}^{S}E_{j}.
\end{align}
\noindent Here, $\sigma_{ij}$ represents the stress tensor, $c_{ijkl}^{E}$ the stiffness tensor at constant electric field, $S_{ij}$ the strain tensor, and $e_{kij}$ the piezoelectric coefficient tensor (with unit C/m$^{2}$). $E_{i}$ is the electric field, $D_{i}$ the electric displacement, and $\varepsilon_{ij}^{S}$ the permittivity tensor at constant strain.

For a lateral electric field applied in the $x$-direction and a piezoelectric medium with tetragonal or hexagonal crystal symmetry (\textit{e.g.}, ScAlN, as considered in the case study below), the induced displacement is aligned with the electric field direction, \textit{i.e.}, along the $x$-axis \cite{42_alma999647780101488}. Furthermore, given the spatial uniformity of the system in the lateral $x$- and $y$-directions, the only non-zero strain component is the $xz$ shear strain, arising from the non-zero displacement gradient in the $z$-direction. Consequently, $ S_{xz} = S_{zx} = \frac{1}{2}\frac{\partial u_{x}}{\partial z}$.

In this case, considering Eq.~\eqref{eq_piezoelectric}, the only non-zero stress component is given by
\begin{equation}
    \sigma_{xz} = \sigma_{zx} = c_{55}^{E} \times 2S_{xz} - e_{15}E_{x},
    \label{eq_piezo_assumption}
\end{equation}
\noindent where $c_{55}^{E}$ is the shear stiffness at constant electric field, $e_{15}$ is the relevant piezoelectric coupling coefficient, and $E_{x}$ is the applied electric field. Subsequently, to simplify the notation, indices will be omitted and replaced by superscripts $^{p}$ for piezoelectric and $^{m}$ for magnetostrictive medium variables and properties.

The elastodynamic equation of motion for the piezoelectric layer can then be written as
\begin{equation}
    \rho^{p}\frac{\partial^{2}u^{p}}{\partial t^{2}} = f_{el}^{p},
    \label{eq_eq_motion}
\end{equation}
\noindent where $\rho^{p}$ is the mass density of the piezoelectric medium, $u^{p} \equiv 
 u_{x}$ is the displacement, and $f_{el}^{p}$ is the elastic body force, given by 
\begin{equation}
	\label{Eq:bodyforce}
     f_{el}^{p} = \frac{\partial}{\partial z} \sigma^{p} = \frac{\partial}{\partial z} (2c^{p}S^{p} - eE) = c^{p}\frac{\partial^{2}u^{p}}{\partial z^{2}}.
\end{equation}
Substituting this expression into the equation of motion \eqref{eq_eq_motion} yields the wave equation
\begin{equation}
     \frac{\partial^{2}u^{p}}{\partial t^{2}} = v^{p}\frac{\partial^{2}u^{p}}{\partial z^{2}},
     \label{eq_resultingmotion}
\end{equation}
\noindent with phase velocity $v^{p} = \sqrt{\frac{c^{p}}{\rho^{p}}}$ and dispersion relation $\omega=v^{p}k$. Mechanical losses within the piezoelectric medium can be incorporated by introducing a complex stiffness constant $c^{p}=c_{r}^{p} + i c_{i}^{p}$. The real part, $c_{r}^{p}$, determines the wavelength and dispersion relation of the acoustic wave, while the imaginary part, $c_{i}^{p}$, accounts for damping. For acoustic excitations, particularly in low-loss FBARs, the imaginary part is typically several orders of magnitude smaller than the real part \cite{42_alma999647780101488}.

The dominant loss mechanism in piezoelectric media at GHz frequencies is typically viscoelastic damping. For this mechanism, the imaginary part of the stiffness can be expressed as \cite{42_alma999647780101488,45_cdi_springer_books_10_1007_978_3_642_24463_6}
\begin{equation}
    c_{i}^{p} = \omega \zeta,
\end{equation}
\noindent where $\omega$ is the angular frequency and $\zeta$ is the viscosity coefficient of the medium. Alternatively, to characterize acoustic resonators, an effective loss factor, the $Q$-factor, is commonly employed. This parameter is more readily accessible experimentally and is related to the complex stiffness constant through \cite{42_alma999647780101488}
\begin{equation}
    Q = \frac{c_{r}^{p}}{c_{i}^{p}}.
\end{equation}
\noindent When losses are included, the wavenumber becomes complex and is given by
\begin{equation}
    k^{p} = \omega\sqrt{\frac{\rho^{p}}{c_{t}^{p}}} = \omega\sqrt{\frac{\rho^{p}}{c_{r}^{p} + c_{i}^{p}}}.
\end{equation}
\noindent Consequently, the phase velocity also becomes complex and can be expressed as $v^{p}=v_{r}^{p} + iv_{i}^{p}$. The square root of a complex number is given by
\begin{equation}
    \sqrt{z} = \sqrt{r}\frac{z+r}{|z+r|},
\end{equation}
\noindent where $r$ is the modulus of the complex number. For $c_{i}^{p} \ll c_{r}^{p}$, the effective complex wavenumber can then be approximated as
\begin{equation}
    k^{p} \approx k_{r}^{p} - ik_{i}^{p},
\end{equation}
\noindent with
\begin{equation}
    k_{i}^{p} = \frac{c_{i}^{p}}{2c_{r}^{p}}k_{r}^{p} = \frac{k_{r}^{p}}{2Q}.
\end{equation}

\subsubsection{Equations of motion in the magnetostrictive film}

\noindent In a magnetostrictive layer, an additional magnetoelastic body force, $f_{mel}$, exists alongside the elastic body force, $f^{m}_{el}$. Consequently, the elastodynamic equation of motion within the magnetostrictive layer becomes
\begin{equation}
    \rho^{m}\frac{\partial^{2}u^{m}}{\partial t^{2}} = f_{el}^{m} + f_{mel}.
    \label{eq_motionMagnet}
\end{equation}
\noindent Analogous to Eq.~\eqref{Eq:bodyforce}, the elastic body force is given by
\begin{equation}
    f_{el}^{m} = \frac{\partial}{\partial z}\sigma^{m} = \frac{\partial}{\partial z}(c_{0}^{m} \times 2S^{m}) = c_{0}^{m}\frac{\partial^{2}u^{m}}{\partial z^{2}},
    \label{eq_fel}
\end{equation}
\noindent where $c_{0}^{m}$ is the stiffness constant of the magnetostrictive material. The magnetoelastic body force is given by (see App.~\ref{App:A})
\begin{equation}
   f_{mel} = 2\frac{B_{1}}{M_{S}^{2}} \begin{bmatrix}
    M_{x}\frac{\partial M_{x}}{\partial x}\\
    M_{y}\frac{\partial M_{y}}{\partial y}\\
    M_{z}\frac{\partial M_{z}}{\partial z}\\
    \end{bmatrix} + \frac{B_{2}}{M_{S}^{2}} \begin{bmatrix}
    M_{x}(\frac{\partial M_{y}}{\partial y} + \frac{\partial M_{z}}{\partial z}) + M_{y}\frac{\partial M_{x}}{\partial y} + M_{z}\frac{\partial M_{x}}{\partial z}\\
    M_{y}(\frac{\partial M_{x}}{\partial x} + \frac{\partial M_{z}}{\partial z}) + M_{x}\frac{\partial M_{y}}{\partial x} + M_{z}\frac{\partial M_{y}}{\partial z}\\
    M_{z}(\frac{\partial M_{x}}{\partial x} + \frac{\partial M_{y}}{\partial y}) + M_{x}\frac{\partial M_{z}}{\partial x} + M_{y}\frac{\partial M_{z}}{\partial y}\\
    \end{bmatrix}.
\end{equation}
\noindent Here, $B_{1}$ and $B_{2}$ represent the magnetoelastic coupling constants of the medium and $M_{S}$ denotes the saturation magnetization. 

Considering only non-zero magnetization gradients along the film thickness (\textit{i.e.}, only $\frac{\partial M_{z}}{\partial z} \neq 0$), we obtain
\begin{equation}
   f_{mel} = 2\frac{B_{1}}{M_{S}^{2}} \begin{bmatrix}
    0\\
    0\\
    M_{z}\frac{\partial M_{z}}{\partial z}\\
    \end{bmatrix} + \frac{B_{2}}{M_{S}^{2}} \begin{bmatrix}
     \frac{\partial M_{z}}{\partial z} + M_{z}\frac{\partial M_{x}}{\partial z}\\
     \frac{\partial M_{z}}{\partial z} + M_{z}\frac{\partial M_{y}}{\partial z}\\
    0\\
    \end{bmatrix}  
\end{equation}
\noindent Furthermore, within the linear magnetic regime, characterized by $M_{y}, M_{z} \ll M_{S}$ and $M_{x} \approx M_{S}$, second-order terms in the magnetization can be neglected. This leads to
\begin{equation}
    f_{mel} \approx \frac{B}{M_{S}}\frac{\partial M_{z}}{\partial z} \hat{e}_{x}
    \label{eq_fmel}
\end{equation}
\noindent with $B \equiv B_{2}.$

Due to the thin nature of the magnetostrictive layer (on the order of the magnetic exchange length), the exchange interaction tends to render magnetic excitations nearly spatially uniform within the layer, \textit{i.e.}, ${\partial M_{z}}/{\partial z} \approx 0$. We note that this approximation has been extensively validated through micromagnetic simulations and is fully consistent with the FEM model described below [see Fig.~\ref{mag_dynam}(b)]. Under these conditions, the magnetoelastic body force becomes negligible, \textit{i.e.}
\begin{equation}
    f_{mel} \approx 0.
    \label{eq_zero_fmel}
\end{equation}

Substituting the body forces, given by Eqs.~\eqref{eq_fel} and \eqref{eq_zero_fmel}, into the equation of motion \eqref{eq_motionMagnet} yields the wave equation
\begin{equation}
    \frac{\partial^{2}u^{m}}{\partial t^{2}} = v_{t}^{m}\frac{\partial^{2}u^{m}}{\partial z^{2}}.
    \label{eq_resultingMotion_magnet}
\end{equation}
\noindent with a phase velocity of \begin{math}v^{m}=\sqrt{\frac{c^{m}}{\rho^{m}}}\end{math}, where \begin{math}c^{m}\end{math} is the intrinsic mechanical stiffness constant and \begin{math}\rho\end{math} the density. The corresponding wavenumber of the magnetostrictive medium is given by $k^{m} = \omega\sqrt{\frac{\rho^{m}}{c^{m}}}$.

Within this approximation, the elastodynamic response can thus be treated as decoupled from the magnetization dynamics. For small-amplitude precessional ferromagnetic resonance (FMR) motion, the dynamics of the magnetization vector $\mathbf{M}$ at angular frequency $\omega$ can be described by the linearized Landau-Lifshitz-Gilbert (LLG) equation (see App.~\ref{App:B})
\begin{equation}
	i\omega \begin{bmatrix}
		M_{y}\\
		M_{z}\\
	\end{bmatrix} = \begin{bmatrix}
		-(\omega_{0}+\omega_{M})M_{z} - 2BS^{m}\gamma\\
		\omega_{0}M_{y}\\
	\end{bmatrix} + i\omega \alpha \begin{bmatrix}
		-M_{z}\\
		M_{y}\\
	\end{bmatrix},
\end{equation}
\noindent with $\gamma$ the gyromagnetic ratio, $\alpha$ the Gilbert damping parameter, and the abbreviations $\omega_{0}=\mu\gamma H_{0}$ and $\omega_{M}=\mu\gamma M_{S}$. $H_0$ denotes an externally applied magnetic field and $\mu$ represents the permeability. The spatially uniform magnetization profile throughout the thickness of the magnetostrictive film can be determined by solving the LLG equation in the long-wavelength limit, corresponding to wavevectors approaching zero ($k\approx 0$). Specifically, this approximation permits the exclusion of the exchange field $\mathbf{H}_{ex} \propto k^2 \mathbf{M}$. 

In the magnetostrictive medium, the spatially-uniform ($k = 0$) magnetization dynamics within the magnetostrictive layer with angular frequency $\omega$ can thus be expressed as a function of the shear strain by (see App.~\ref{App:B})
\begin{equation}
	M_{z} = \frac{2B\gamma\omega_{y}}{\omega^{2}-\omega_{y}\omega_{z}}S^{m}_0,
	\label{eq_Mz}
\end{equation}
\noindent with, $\omega_{y}=\omega_{0}+i\omega \alpha$, $\omega_{z} = \omega_{0} + \omega_{M} + i\omega\alpha$, and $S^{m}_0 \equiv S^{m}(k = 0)$. This approximation will be employed to evaluate the power transfer from the elastic to the magnetic domain in Sec.~\ref{sec:transduction_eff}.

We note that an analytical model incorporating backaction from the magnetic to the elastic system can be derived without  Eq.~\eqref{eq_zero_fmel}, \textit{e.g.}, within a linear response approximation. Such a model is presented in \cite{vanderveken2022powertransfermagnetoelectricresonators} and yields quantitative results that are not significantly different from those obtained using the model presented here. Nonetheless, micromagnetic simulations as well as the FEM simulations described in Sec.~\ref{Sec:FEM} indicate that it is difficult to design magnetoelectric FBARs, for which Eq.~\eqref{eq_zero_fmel} is strongly violated. Consequently, the model presented in this work is expected to be applicable to most FBAR geometries and material systems.

\subsubsection{Solution of the boundary value problem}

\noindent As derived above, the dynamic response of the displacement in both the piezoelectric and magnetostrictive layers is governed by wave equations. Consequently, a wave-based ansatz can be employed as a solution to the equations of motion Eq.~\eqref{eq_resultingmotion} and Eq.~\eqref{eq_resultingMotion_magnet}. However, due to differing phase velocities, the wave numbers in each layer vary at a given frequency. Within the piezoelectric medium, the displacement and strain ansatzes can be expressed as
\begin{align}
    u^{p}(z) &= A_{1}e^{ik^{p}z} + A_{2}e^{-ik^{p}z} \\
    S^{p}(z) &= \frac{ik^{p}}{2}(A_{1}e^{ik^{p}z} - A_{2}e^{-ik^{p}z})
    \label{eq:strain_p}
\end{align}
\noindent Following Eq.~\eqref{eq_piezo_assumption}, the total stress within the piezoelectric layer becomes
\begin{equation}
    \sigma^{p}(z) = 2c^{p}S^{p}(z) - eE = ic_{t}^{p}k^{p} (A_{1}e^{ik^{p}z} - A_{2}e^{-ik^{p}z}) - eE
\end{equation}
\noindent Within the magnetostrictive layer, the displacement and strain ansatzes are given by
\begin{align}
    u^{m}(z) &= A_{3}e^{ik^{m}z} + A_{4}e^{-ik^{m}z} \\
    S^{m}(z) &= \frac{ik^{m}}{2}(A_{3}e^{ik^{m}z} - A_{4}e^{-ik^{m}z})
    \label{eq:strain_m}
\end{align}
\noindent The total stress within the magnetostrictive layer becomes
\begin{equation}
    \sigma^{m}(z) = ic^{m}k^{m} (A_{3}e^{ik^{m}z} - A_{4}e^{-ik^{m}z}).
\end{equation}

The boundary conditions impose zero stress at the top surface, continuity of displacement and stress at the piezoelectric--magnetostrictive interface, and zero stress at the bottom surface, \textit{i.e.}
\begin{align}
    \sigma^{p}(z=d) &= 0 \\
    \sigma^{p}(z=0) &= \sigma^{m}(z=0) \\
    u^{p}(z=0) &= u^{m}(z=0) \\
    \sigma^{p}(z=-t) &= 0.
\end{align}
\noindent These four conditions enable the determination of the four amplitudes $A_{1}$ to $A_{4}$. Substituting these expressions into Eqs.~\eqref{eq:strain_p} and \eqref{eq:strain_m} yields expressions for the strain within the device. In the piezoelectric layer, the strain is given by
\begin{align}
    S^{p}(z) &= eE \frac{c^{m}k^{m}\sin ( k^{m}t )\cos ( k_{t}^{p}z )+2c^{p}k_{t}^{p}\sin\!\left(\frac{1}{2}k_{t}^{p}d\right)\cos ( k^{m}t )\cos\!\left(\frac{1}{2}k_{t}^{p}\left( d-2z\right)\right)}{2c^{m}c_{t}^{p}k^{m}\cos (k_{t}^{p}d )\sin (k^{m}t )+2\left(c_{t}^{p}\right)^{2}k_{t}^{p}\sin (k_{t}^{p}d )\cos (k^{m}t )} \nonumber \\
    &\equiv eES^{p}_{k},
\end{align}
\noindent whereas in the magnetostrictive layer, the strain is given by
\begin{equation}
    S^{m}(z)=eE\frac{k^{m}\sin^{2}\!\left(\frac{1}{2}k_{t}^{p}d\right)\sin\!\left(k^{m}(t+z)\right)}{c^{m}k^{m}\cos\!\left(k_{t}^{p}d\right)\sin\!\left(k^{m}t\right) + c_{t}^{p}k_{t}^{p}\sin\!\left(dk_{t}^{p}\right)\cos\!\left(k^{m}t\right)}.
    \label{eq:Strain_M}
\end{equation}
\noindent Note that the strain in both layers is a complex quantity, owing to the complex stiffness and wavenumber in the piezoelectric layer within this model.

Equation \eqref{eq:Strain_M} can be further simplified under the condition that the magnetic layer is much thinner than the acoustic wavelength (\textit{i.e.}, $k^{m}(t+z)\ll 1$). This leads to
\begin{equation}
	S^{m}(z) \approx eE(t+z)\frac{\left(k^{m}\right)^{2}\sin^{2}\!\left(\frac{1}{2}k_{t}^{p}d\right)}{D(\omega)} \equiv eE(t+z)S_{k}^{m},
	\label{approx_strain_magnet}
\end{equation}
\noindent with the strain denominator $D(\omega)$ given by
\begin{equation}
	D(\omega) = c^{m}k^{m}\cos\!\left(k_{t}^{p}d\right)\sin\!\left(k^{m}t\right) + c_{t}^{p}k_{t}^{p}\sin\!\left(k_{t}^{p}d\right)\cos\!\left(k^{m}t\right).
	\label{eq_StrainDenominator}
\end{equation}

Substitution of Eq.~\eqref{eq:Strain_M} into Eq.~\eqref{eq_Mz} yields the magnetization dynamics within the magnetostrictive layer that are driven by the dynamic strain field. Consistent with the preceding analysis of linear magnetization dynamics, the long-wavelength limit, characterized by a wavevector approaching zero ($k=0$), this dynamic strain field can be approximated as an effective spatially uniform strain. This uniform strain $S^{m}_0$ generates a spatially uniform magnetization precession profile that is equivalent to that induced by the complete, spatially varying strain distribution. The derivation of this effective strain in the long-wavelength regime, specifically for the case of vanishing wavevector (k=0), is presented in detail in App.~\ref{App:C}, leading to a magnetization precession amplitude of
\begin{equation}
	M_{z} = \frac{2B\gamma\omega_{y}eE}{ t(\omega^{2}-\omega_{y}\omega_{z})} \frac{k^{m}\sin^{2}\!\left(\frac{1}{2}k_{t}^{p}d\right)\left(k^{m} - k^{m}\cos\!\left(k^{m}t\right)\right)}{(c^{m}k^{m}\cos\!\left(k_{t}^{p}d\right)\sin\!\left(k^{m}t\right) + c_{t}^{p}k_{t}^{p}\sin\!\left(dk_{t}^{p}\right)\cos\!\left(k^{m}t\right))(k^{m})^{2}}.
	\label{Mz_model}
\end{equation} 

\section{Finite Element Simulations of Magnetoelectric FBARs \label{Sec:FEM}}

\noindent In this section, we introduce a numerical model of the magnetoelectric resonator, implemented within the finite-element multiphysics software COMSOL\texttrademark. Extending the analytical model, this numerical approach enables a comprehensive description of the coupling between the magnetic and elastic systems without the need for approximations. In particular, it incorporates the nonlinear Landau--Lifshitz--Gilbert (LLG) equation, including the exchange interaction, to compute the magnetization dynamics. Moreover, it takes the potential effect of magnetostrictive backaction into account, describing the detailed interactions between elastic and magnetic subsystems. By comparing the results obtained from the analytical and numerical models, we can then validate the assumptions underlying the analytical model. Furthermore, the FEM model facilitates an investigation of the (nonlinear) resonator dynamics, permitting the identification of transient regimes and the determination of resonant frequencies.

\subsection{Model and parameter description}

\begin{figure}[p] 
	\centering
	\includegraphics[scale=0.7]{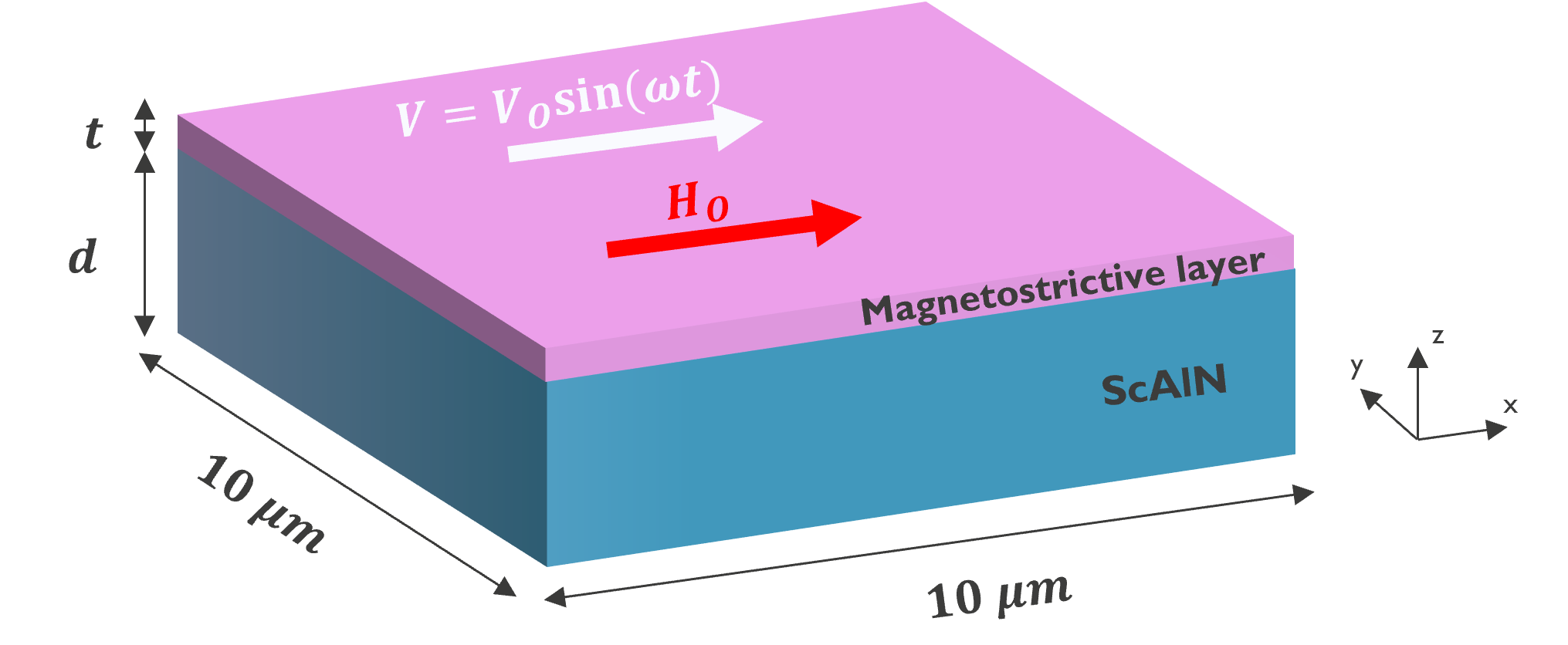}
	\caption{FEM representation of the magnetoelectric resonator. An oscillating electric field with amplitude $V_0$ and angular frequency $\omega$ is applied across the piezoelectric ScAlN layer to excite standing elastic shear waves, which generate magnetization dynamics in the magnetostrictive layer. $H_0$ denotes the static external magnetic field, as in the analytical model.}
	\label{FEM_model}
\end{figure}

\noindent A schematic illustration of the FEM model of the magnetoelectric resonator is presented in Fig.~\ref{FEM_model}. To approximate a laterally infinite structure, the in-plane dimensions of the model were chosen to be substantially larger than the total thickness, minimizing edge effects and emulating translational symmetry in the lateral directions. Specifically, the in-plane lateral dimensions were set to 10 $\mu$m, while the thicknesses of the piezoelectric (with ScAlN parameters) and magnetostrictive layers were defined as $d = 200$~nm and $t = 20$~nm, respectively. Consistent with the assumptions of the analytical model, the resonator was treated as laterally infinite, with only the shear strain component in the $xz$-plane considered non-zero. This simplification was realized in the FEM framework by employing a one-dimensional mesh along the $z$-axis and imposing displacement boundary conditions $u_{y} = u_{z} = 0$ at the lateral edges. These constraints ensured uniformity of both strain and magnetization in the in-plane directions, confining all spatial variation to the out-of-plane $z$-direction.

Under the assumption of the magnetization constrained varying only along the out-of-plane $z$-axis, the demagnetizing field was approximated as $H_\mathrm{demag} = -M_{S}m_{z}$. Viscoelastic losses within the piezoelectric layer were incorporated via a frequency-dependent viscous damping model. Specifically, the mechanical damping was characterized using a Rayleigh-type formulation, wherein the stiffness matrix was scaled by a damping coefficient $\beta = 1/\left(\omega Q\right)$ \cite{RayleighDamping}, with $\omega$ representing the angular frequency and $Q$ the resonator's mechanical $Q$-factor. This approach accounts for the frequency-dependent nature of energy dissipation and enables the modulation of damping characteristics within the piezoelectric layer to emulate a broad range of mechanical $Q$-factor values.

\subsection{Dynamic response of the magnetoelectric resonator}

\begin{figure}[p] 
	\centering
	\includegraphics[scale=0.7]{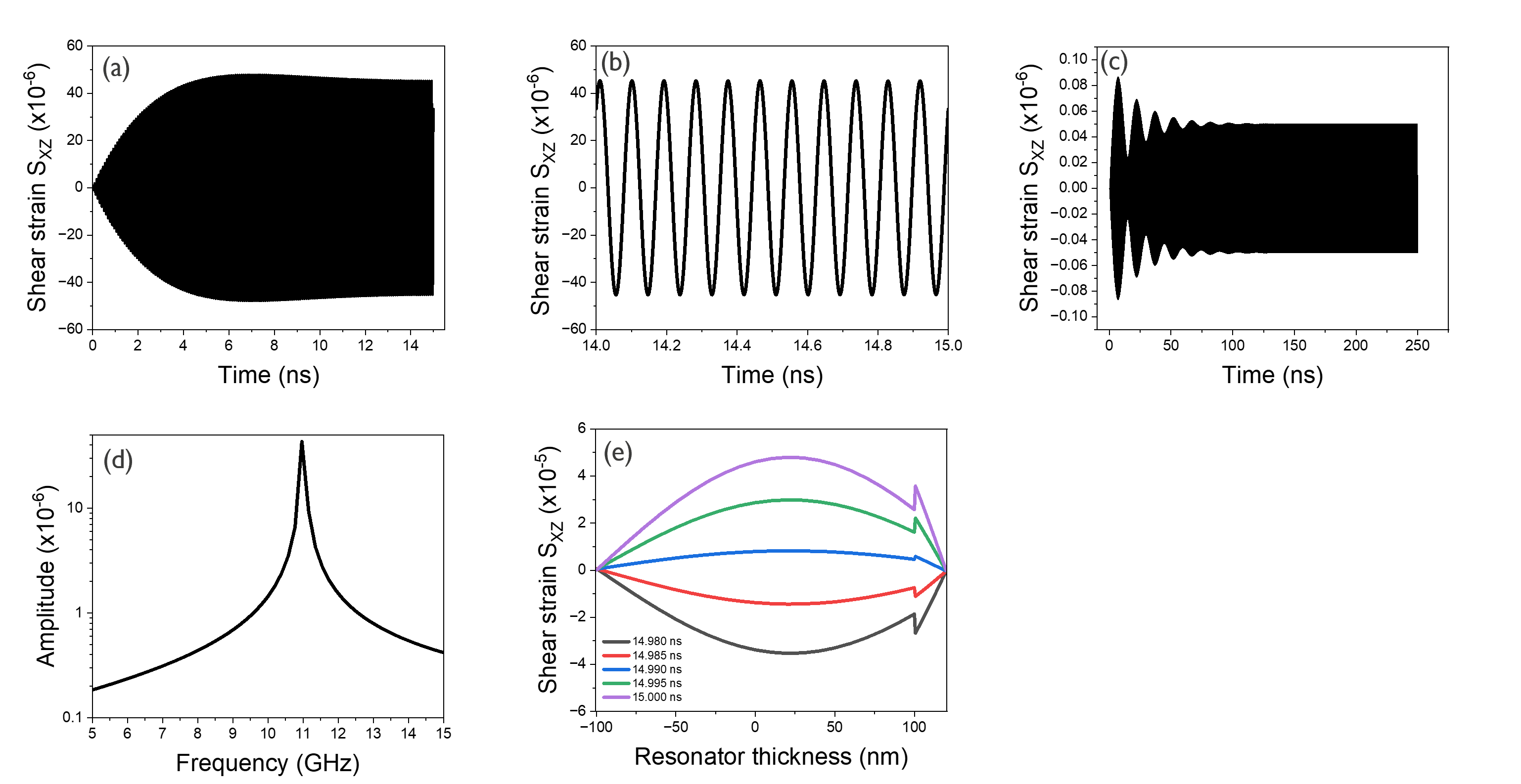}
	\caption{ Shear strain dynamics in a ScAlN/CoFeB magnetoelectric resonator under varying mechanical $Q$-factors and excitation voltages. (a) Time evolution of the shear strain for $Q = 100$ and excitation voltage $V_0 = 1$ V, illustrating the transient buildup and saturation of the shear strain amplitude. (b) Magnified view of the shear strain waveform from 14 to 15 ns in (a), highlighting steady-state oscillations. (c) Strain dynamics for $Q = 1000$ and $V_0 = 1$ mV, showing an extended transient regime and lower strain amplitude due to reduced excitation. (d) Fast Fourier transform (FFT) spectrum of the steady-state regime in (a), revealing a single dominant frequency component. (e) Shear strain profiles across the resonator thickness at selected time instances for the parameters in (a), illustrating the spatial mode profile of the elastic resonance.
	\label{Time_behavo}}
\end{figure}

In the numerical simulations, a sinusoidal voltage, $V(t) = V_{0}\sin\left(\omega t\right)$, with an angular frequency $\omega = 2\pi \nu$, was applied across the resonator along the $x$-direction, as schematically depicted in Fig.~\ref{FEM_model}. This excitation generated mechanical oscillations within the structure, specifically dynamic strain fields. The time evolution of the shear strain amplitude, shown in Figs.~\ref{Time_behavo}(a) and \ref{Time_behavo}(c) for mechanical $Q$-factors of 100 and 1000, respectively, illustrates the system's progression from an initial transient response to a steady-state regime. To provide a clearer view of the oscillatory behavior, a magnified 1 ns time window of the shear strain response from Fig.~\ref{Time_behavo}(a) is presented in Fig.~\ref{Time_behavo}(b). The steady-state regime is characterized by periodic shear strain oscillations with constant amplitude, indicative of sustained resonant behavior. As anticipated for underdamped harmonic systems, an increase in the $Q$-factor leads to a longer transient before reaching steady-state oscillations.

We note that the excitation voltage amplitude $V_0$ used in the simulations in Fig.~\ref{Time_behavo}(c) was reduced by a factor of 1000 relative to the amplitude employed in Fig.~\ref{Time_behavo}(a). This deliberate scaling of the driving voltage led to a corresponding decrease in the steady-state strain amplitude, despite the resonator exhibiting a higher mechanical $Q$-factor. In all simulations, the amplitude of the applied excitation voltage was carefully selected to ensure operation within the linear response regime of the resonator to avoid nonlinear effects. To empirically verify the linearity of the system’s response, a time-domain Fast Fourier Transform (FFT) was performed on the shear strain signal obtained from the simulation with $Q=100$ [see Fig.~\ref{Time_behavo}(a)]. The resulting frequency spectrum, shown in Fig.~\ref{Time_behavo}(d), revealed a single dominant spectral component centered at 11 GHz, corresponding precisely to the excitation frequency. The absence of significant higher-order harmonic or subharmonic peaks in the spectrum confirms the linear nature of the resonator’s mechanical response under the applied excitation conditions.

Figure~\ref{Time_behavo}(e) presents the spatial distribution of the shear strain along the thickness of the resonator at five discrete time instances, providing further insight into the dynamic behavior of the FEM model. The strain profiles clearly exhibit characteristic first-order resonant mode shapes, in keeping with the analytical model. Notably, a discontinuity in the shear strain field is observed at the interface between the piezoelectric and magnetostrictive layers. This interfacial discontinuity arises from the mechanical impedance mismatch between the two materials and is in agreement with the analytical model discussed in the preceding section.

\begin{figure}[p] 
	\centering
	\includegraphics[scale=0.7]{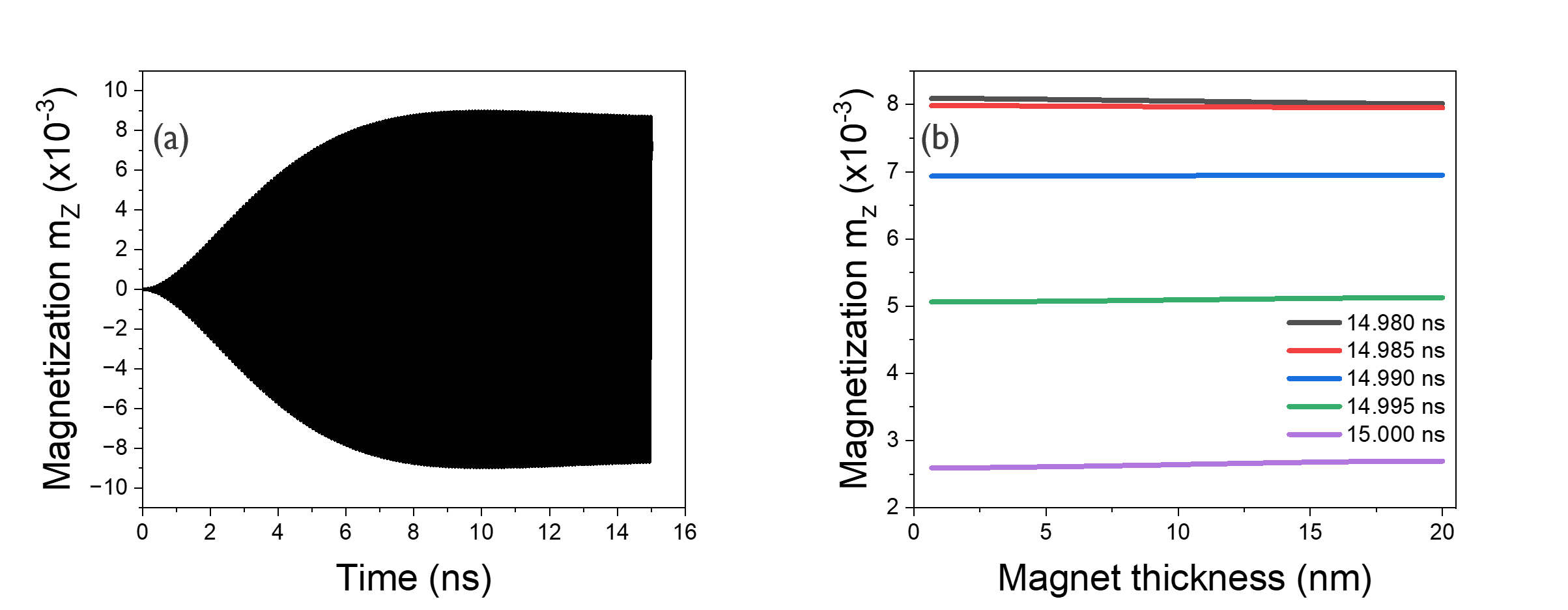}
	\caption{Magnetization dynamics in a ScAlN/CoFeB-based magnetoelectric resonator with a mechanical quality factor $Q = 100$ and excitation amplitude $V_0 = 1$ V. (a) Time evolution of the spatially averaged out-of-plane magnetization component $m_z$. (b) Spatial profiles of $m_z$ across the magnetostrictive layer thickness at selected time instants, highlighting the spatial uniformity of the magnetic response under shear strain excitation.}
	\label{mag_dynam}
\end{figure}

The standing shear strain waves depicted in Fig.~\ref{Time_behavo}(e) couple to the magnetization within the magnetostrictive layer and induce magnetization dynamics. This is illustrated in Fig.~\ref{mag_dynam}(a), which shows the temporal evolution of the normalized out-of-plane magnetization component, $m_z = M_z/M_S$. To verify that the observed magnetization dynamics correspond to an FMR mode, the externally applied magnetic field was systematically varied. As discussed above, the spatial profiles of the magnetization precession amplitude across the thickness of the magnetostrictive layer in Fig.~\ref{mag_dynam}(b) are nearly uniform. This substantiates the assumption of spatially homogeneous magnetization dynamics, which underpins the analytical model developed in the preceding sections. 

\section{Magnetoelectric transduction efficiency\label{sec:transduction_eff}}

\noindent Drawing upon both the analytical and FEM models, we can now quantify the magnetic and elastic power dissipation and energy transfer within the resonator under steady-state conditions and assess the transducer efficiency. Subsequently, in Sec.~\ref{sec:case_study}, we will use these efficiency calculations to investigate the impact of material parameters and geometric dimensions on the performance of the magnetoelectric resonator.

\subsection{Magnetic power transfer}

\noindent The strain field within the magnetostrictive layer induces a magnetoelastic effective magnetic field, which serves as the driving force for the excitation of magnetization dynamics. As a consequence, a fraction of the total energy input into the resonator is transferred into the magnetic subsystem. The power density in the magnetic subsystem, originating from the magnetoelastic interaction, can be expressed by \cite{46_KobayashiT.1973FRiT,47_WeilerM.2011Edfr}
\begin{equation}
    p(z) = \frac{i\omega}{2}\mu_{0}\mathbf{M}_{dyn}\cdot \mathbf{H}_{mel}^{*}.
\end{equation}
\noindent The power density can be decomposed into real and imaginary components, \textit{i.e.}, $p(z) \equiv p_{r}(z) + ip_{i}(z)$. The real component, $p_{r}$, corresponds to the energy dissipation within the magnetic material, whereas the imaginary component, $p_{i}$, represents the oscillatory reactive power. Integrating the real part of the power density over the entire volume of the magnetic layer yields the total power absorbed by the magnetic subsystem. Consequently, the power absorption per unit area in a one-dimensional model is given by the spatial integral of $p_{r}(z)$ over the thickness of the magnetostrictive layer
\begin{equation}
	P_{m} = \int_{-t}^{0} \!p_{r}(z)\,dz.
	\label{eq_magnetic_power}
\end{equation}

Within the framework of the FEM model, the magnetic power density can be directly evaluated through numerical integration of Eq.~\eqref{eq_magnetic_power} over the computed $z$-profiles of the magnetization and the magnetoelastic field vectors. Conversely, in the analytical model, with the spatial distribution of the magnetization defined by Eq.~\eqref{eq_Mz} and the magnetoelastic field in Eq.~\eqref{eq_resultingHmel}, the expression for the power density transferred into the magnetic system assumes the following form:
\begin{align}
    p(z) & =  i\frac{\omega}{2}\mu_{0} \left(\frac{2B\gamma\omega_{y}}{\omega^{2} - \omega_{y}\omega_{z}}S^{m}_0\right)\left(\frac{-2B(S^{m})^{*}}{\mu_{0}M_{S}}\right) \nonumber \\
    & = -i\frac{2\omega B^{2}\gamma \omega^{y}}{M_{S}\left( \omega^{2}-\omega_{y}\omega_{z}\right) } S^{m}_0\, S^{m}(z) \nonumber \\
    & = -i2\omega c_{B}^{m} S^{m}_0\, S^{m}(z),
\end{align}
\noindent with $c_{B}^{m} = \frac{B^{2}\gamma\omega_{y}}{M_{S}\left( \omega^{2}-\omega_{y}\omega_{z}\right)}$. 
Using Eq.~\eqref{approx_strain_magnet} and performing the integration leads to the total absorbed magnetic power given by
\begin{align}
	P_{m} & = 2\omega c_{B}^{m} S^{m}_0 \int_{-t}^{0} \!S_{r}^{m}(z)\,dz.
	\label{eq_magnetic_power_analyt} \\
& = \omega c_{B}^{m} S^{m}_0 eEt^{2}S^{m}_{k}(\omega).
	\label{total_magnetic_power}
\end{align}

Equation \eqref{total_magnetic_power} offers several insights into the magnetic power absorption in a piezoelectric--magnetostrictive bilayer resonator. First, as shown by Eq.~\eqref{eqC3:Strain_M}, $S^{m}_0$ is linearly dependent on the applied electric field $E$, and thus the magnetic power absorption exhibits a quadratic dependence on $E$. Consequently, increasing the electric field leads to rapidly increasing magnetization dynamics, until nonlinear effects become dominant. Moreover, $S^{m}_0$ scales linearly with the piezoelectric constant $e \equiv e_{15}$. Therefore, also the power absorption scales quadratically with $e_{15}$. Additionally, Eq.~\eqref{total_magnetic_power} reveals a quadratic dependence on the magnetoelastic coupling constant $B$, implying that resonators employing magnetostrictive materials with larger $B$ values are expected to exhibit higher transduction efficiencies. By contrast, the influence of other magnetic material parameters, such as the saturation magnetization $M_S$ or the Gilbert damping $\alpha$, is less pronounced.

From a geometric standpoint, the power absorption is strongly influenced by the relative thicknesses of the magnetic and piezoelectric layers. However, the thicknesses also indirectly affect the strain amplitude, making it challenging to formulate a simple general relationship between thickness and magnetic power absorption. In Sec.~\ref{sec:case_study}, we discuss a specific case study using different sets of material constants to examine the impact of material and geometric parameters on power absorption in both the mechanical and magnetic subsystems.

The maximum power transfer into the magnetic domain occurs when the elastic resonance frequency coincides with the ferromagnetic resonance frequency. While tuning the elastic resonance frequency necessitates modifications to the structural dimensions of the resonator, the magnetic resonance frequency can be adjusted by varying an external magnetic field $H_0$. Below, we derive an approximate expression for the external field required to achieve simultaneous resonance in both the elastic and magnetic systems. This expression can, \textit{e.g.}, serve as a guideline for optimizing experimental conditions.

In the limit of zero losses, the strain denominator in Eq.~\eqref{eq_StrainDenominator} vanishes at resonance, leading to an infinite strain amplitude. However, the introduction of viscoelastic losses results in a complex-valued denominator. At resonance, the real part of the denominator becomes zero, while the finite imaginary part limits the strain amplitude. To determine the resonance frequency, the real part of the denominator is equated to zero, and the resulting equation is solved for the frequency. For a bilayer system, solving this equation analytically can be challenging, necessitating further simplifications. Typically, the imaginary components of the complex stiffness and wavenumber are significantly smaller than their real counterparts, allowing us to neglect their influence on the resonance frequency. Thus, we can approximate $c_{i}^{p}\approx 0$. This simplification leads to the following resonance condition:
\begin{equation}
    c^{m}k^{m}\cos (k_{r}^{p}d)\sin (k^{m}t) + c_{r}^{p}k_{r}^{p}\sin (k_{r}^{p}d )\cos (k^{m}t )=0.
\end{equation}
\noindent For a thin magnetostrictive layer, where $t \ll d$, the approximate solution to this resonance condition, corresponding to the elastic resonance frequency, is given by
\begin{equation}
    \nu_{res}^{el} \approx \frac{nv_{r}^{p}}{2\left( d+t\right)},
\end{equation}
\noindent with $n$ the mode number. This corresponds to the mechanical resonance frequency of a single piezoelectric layer of thickness $d + t$.

Maximum power transfer is achieved when the ferromagnetic resonance frequency, tuned by the external magnetic field, aligns with the mechanical resonance frequency. The condition for simultaneous resonance is given by
\begin{equation}
    \nu_{res}^{el}=\nu_{res}^{m} \Rightarrow \frac{nv_{r}^{p}}{2\left( d+t\right)} = \frac{\sqrt{\omega_{0}\left(\omega_{0} + \omega_{M}\right)}}{2\pi}.
    \label{eq_resonance_req}
\end{equation}
\noindent The magnetic field required to achieve a ferromagnetic resonance frequency that satisfies Eq.~\eqref{eq_resonance_req} is given by
\begin{equation}
    H_{0} = \frac{-\omega_{M}+\sqrt{\omega_{M}^{2}+ 4(\frac{n\pi v_{r}^{p}}{d+t})^{2}}}{2\mu \gamma}.
    \label{eq:H0}
\end{equation}
\noindent This indicates that bilayer structures incorporating piezoelectric materials with higher shear wave group velocities require larger external magnetic fields to achieve simultaneous mechanical and ferromagnetic resonance.

\subsection{Elastic power loss}

\noindent In the preceding section, analytical expressions were derived to quantify the power transfer associated with the excitation of magnetization dynamics within the magnetostrictive layer, driven by the dynamic mechanical displacement. In the following, we address the viscoelastic power dissipation within the resonator. The general expression for the elastic power density is given by
\begin{equation}
    p_{el} = \frac{i\omega}{2}\sum_{i,j,k,l} S_{ij}(c_{ijkl}S_{kl})^{\star}.
\end{equation}
\noindent In the geometry considered in this work, only the shear strain component $S_{xz}=S_{zx}$ contributes to the mechanical energy dissipation in the resonator. For materials exhibiting tetragonal or hexagonal crystal symmetry, the power density associated with this shear deformation can be expressed as
\begin{equation}
    p_{el}(z)=i\omega (c_{t}^{p*}|S^{p}(z)|^{2} + c_{0}^{m}|S^{m}(z)|^{2}).
\end{equation}
\noindent Similar to the magnetic power, the power per unit area can be obtained by integrating over the resonator thickness, leading to 
\begin{align}
    P_{el} &= i\omega \left( c_{t}^{p*} \int_{0}^{d} \!|S^{p}(z)|^{2}\,dz + c_{0}^{m}\int_{-t}^{0} \!|S^{m}(z)|^{2}\,dz \right)\\
    &\approx i\omega c_{t}^{p*} \int_{0}^{d} \!|S^{p}(z)|^{2}\,dz    \label{eq_approx_elastic} \\
    &= i\omega \left( c_{r}^{p} - ic_{i}^{p}\right) e^{2}E^{2} \int_{0}^{d} \!|S_{k}^{p}(z)|^{2}\,dz.
\end{align}

The approximation in Eq.~(\ref{eq_approx_elastic}) holds under the condition that the magnetostrictive layer is significantly thinner than the piezoelectric layer, as consistently assumed throughout this work. The real part of this expression corresponds to the elastic power dissipation by viscoelastic effects and is given by
\begin{equation}
    P_{el} = \omega E^{2}\frac{c_{r}^{p}e^{2}}{Q} \int_{0}^{d} \!|S_{k}^{p}(z)|^{2}\,dz.
    \label{eq_elastic_power}
\end{equation}
\noindent Analogous to magnetic power absorption, the elastic power loss exhibits a quadratic dependence on both the applied electric field and the piezoelectric coefficient, \textit{i.e}., $P_{el} \propto e^{2}E^{2}$. However, as the resulting strain amplitude is also influenced by additional parameters, a straightforward analytical relationship between the applied field and the dissipated power cannot be established. Consequently, numerical calculations become essential for accurately capturing the interrelations between these factors and for systematically exploring the effects of material parameters and structural dimensions on energy dissipation.

\subsection{Magnetoelectric power transfer efficiency}

When employed as a magnetoelectric transducer to excite magnetization dynamics from electrical signals, a key performance metric is the power transfer efficiency into the magnetic subsystem. The total power supplied to the system can be decomposed into three primary components: reactive power, elastic power loss, and magnetic power absorption. The reactive power corresponds to the non-dissipative energy that oscillates within the transducer during resonance and contributes only during the transient regime, prior to reaching steady-state conditions. In contrast, the steady-state power transfer efficiency is governed by the balance between magnetic and elastic dissipation. In an ideal magnetoelectric transducer, energy losses occur predominantly within the magnetic subsystem, while elastic losses are considered parasitic and should be minimized to optimize performance. We note that, In the system considered here, elastic energy dissipation occurs primarily within the piezoelectric layer because of its substantially greater volume relative to the magnetostrictive layer.

By quantitatively comparing the mechanical (elastic) and magnetic power dissipation components, the efficiency of magnetic transduction in the magnetoelectric resonator can be defined as
\begin{equation}
    \eta = \frac{P_{m}}{P_{m} + P_{el}} .
    \label{eq:eff}
\end{equation}
\noindent 

Here, $P_{m}$ and $P_{el}$ are defined by Eqs.~\eqref{total_magnetic_power} and \eqref{eq_elastic_power}, respectively. The transduction efficiency quantifies the fraction of total active power within the resonator that is effectively converted into magnetic excitations. It is important to emphasize that this efficiency metric, as well as its dependence on material properties and geometric parameters, may differ significantly from the absolute magnitude of power dissipated in the magnetic subsystem.

Equations~\eqref{total_magnetic_power}, \eqref{eq_elastic_power}, and \eqref{eq:eff} reveal that the transduction efficiency is independent of the resonator area, the piezoelectric coefficient $e$, and the magnitude of the applied electric field $E$. This observation leads to several notable implications. First, the efficiency is independent of the transducer's lateral dimensions, which indicates favorable scaling behavior of magnetoelectric resonators. Second, decreasing the applied electric field or voltage affects only the absolute magnitude of power dissipation, without impacting the energy transfer efficiency into the magnetic subsystem. Perhaps more unexpectedly, the piezoelectric coefficient $e$, which governs the strength of electromechanical coupling, also has no impact on the efficiency. This implies that the dominant factors governing transduction efficiency are the viscoelastic losses and the mechanical quality factor $Q$, whereas the applied electric field and piezoelectric constant primarily influence the total power absorption.

These analytical findings are further supported by finite element method (FEM) simulations. In one set of simulations conducted for CoFeB with a mechanical $Q$-factor of 100, varying the applied voltage showed that a lower voltage led to reduced total power dissipation---affecting both magnetic and elastic losses---while the transduction efficiency remained constant. In a separate simulation, reducing the piezoelectric constant $e$ likewise resulted in a lower total power loss, again without affecting the efficiency. These results confirm that, while the absolute power consumption depends on the excitation strength and material parameters, the efficiency of magnetoelectric energy conversion is governed primarily by mechanical dissipation mechanisms. A detailed analysis of the influence of viscoelastic losses and the mechanical $Q$-factor on transduction efficiency is presented in Sec.~\ref{sec:case_study}.

\section{Comparative analysis of S{\MakeLowercase c}A{\MakeLowercase l}N resonators with different magnetostrictive materials \label{sec:case_study}}

To obtain deeper quantitative insights into the performance of magnetoelectric resonators, we present a comprehensive case study utilizing both the analytical model and finite element method (FEM) simulations. This approach enables a systematic investigation of the influence of key geometric and material parameters on the transducer efficiency. Moreover, it provides a means to evaluate the accuracy and applicability of the analytical model by benchmarking its predictions against the results obtained from FEM simulations.

The case study focuses on a magnetoelectric resonator employing hexagonal ScAlN as the base piezoelectric material \cite{48_FichtnerSimon2019AAIs,49_PetrichR.2019IoSf,50_WangJialin2020AFBA}. The relevant physical properties of ScAlN are summarized in Table~\ref{table:1}. The thickness of the piezoelectric layer was fixed at $d = 200$ nm. Unless otherwise stated, the mechanical quality factor of the piezoelectric layer was assumed to be $Q=1000$. In the analytical model, viscoelastic losses were incorporated by introducing an imaginary component to the piezoelectric stiffness, calculated as $c_{i}^{p} = c_{r}^{p}/Q$ \cite{52_ParkMingyo2019SHSA}. In contrast, the FEM simulations accounted for damping effects via a viscous damping coefficient defined as $\beta = 1/\left(\omega Q\right)$, as detailed in the previous section \cite{RayleighDamping}. In addition to ScAlN, three distinct magnetostrictive materials were considered to evaluate their influence on transduction efficiency: CoFeB, Ni, and Terfenol-D (Tb$_{0.3}$Dy$_{0.7}$Fe$_2$). The material properties for these magnetostrictive layers were extracted from the literature and are likewise compiled in Table~\ref{table:1}.

As previously discussed, the ferromagnetic resonance (FMR) frequency can be tuned by varying the externally applied magnetic field $H_0$, as approximated by Eq.\eqref{eq:H0}. In the subsequent analysis, $H_0$ is adjusted for each magnetostrictive material to ensure that the magnetic resonance frequency is matched to the mechanical resonance frequency of the resonator. This resonance alignment maximizes the efficiency of elastic-to-magnetic energy transfer by optimizing the magnetoelastic coupling. The values of the applied magnetic fields and the corresponding resonance frequencies used in both the analytical and FEM models are summarized in Table~\ref{table:2}.

In the following, we systematically investigate the influence of key geometric and material parameters on the transduction efficiency of the magnetoelectric resonator. Particular attention is given to the role of mechanical energy dissipation, as characterized by the mechanical quality factor $Q$, which governs viscoelastic losses within the resonator.

\subsubsection{Influence of the magnetostrictive layer thickness}

The dimensional parameters of the resonator significantly influence the strain amplitude, thereby affecting power transfer efficiency and overall transducer performance. To investigate the impact of these parameters, we have computed both the magnetic and elastic power losses, as well as the transducer efficiency, for the three selected magnetostrictive materials, considering mechanical quality factors $Q = 100$ and $Q = 1000$, across a range of magnetostrictive layer thicknesses, $t$. It is important to note that, for each thickness, the external magnetic field was tuned to align the magnetic resonance frequency with the mechanical resonance frequency, thereby ensuring optimal magnetoelastic coupling.

\begin{figure}[p] 
	\centering
	\includegraphics[scale=0.7]{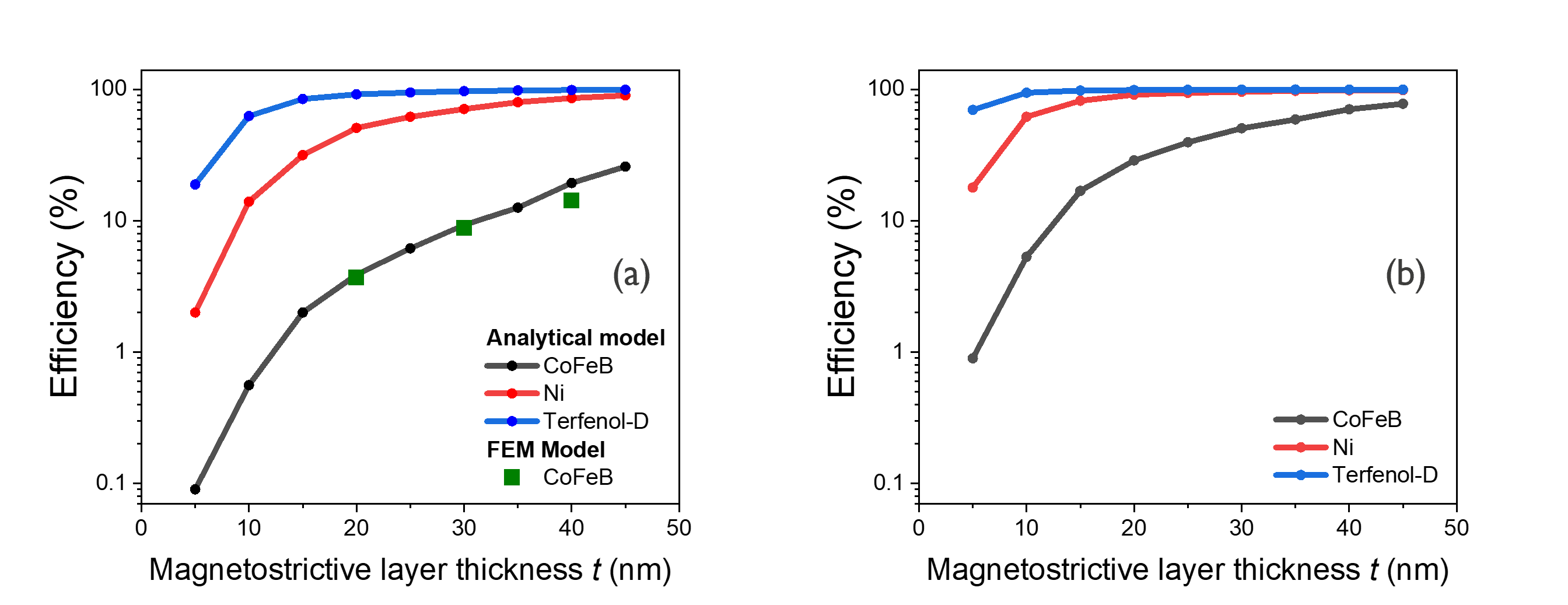}
	\caption{Magnetic transduction efficiency $\eta$ as a function of the magnetostrictive layer thickness $t$ for ScAlN resonators incorporating CoFeB, Ni, and Terfenol-D material parameters, evaluated for (a) $Q = 100$ and (b) $Q = 1000$. In each case, the externally applied magnetic field was tuned to ensure frequency matching between the elastic and ferromagnetic resonances.}
	\label{eff_ifo_t}
\end{figure}

Figure~\ref{eff_ifo_t} demonstrates that, for thin magnetostrictive layers, the transduction efficiency is relatively low. This reduced efficiency arises from low magnetic power absorption, as the small magnetostrictive volume limits the magnetic losses. As the thickness of the magnetic layer increases, the magnetostrictive (magnetic) volume expands, resulting in higher magnetic power absorption and, consequently, improved efficiency.

The transduction efficiency is a critical performance metric when magnetoelectric transducers are employed for magnetic excitation. For all materials considered, the efficiency increases with magnetostrictive layer thickness $t$, driven by the corresponding rise in magnetic power dissipation. Notably, for the selected materials, particularly Ni and Terfenol-D, remarkably high magnetic transduction efficiencies, approaching 100\%, are achievable. In contrast, while CoFeB exhibits slightly lower efficiencies, it still attains values exceeding 70\% for magnetostrictive layers as thin as 40 nm at a mechanical $Q$-factor of 1000. These findings confirm the expectation that magnetoelectric transducers can efficiently excite FMR or spin waves, with magnetic losses becoming dominant over elastic losses within the parameter space explored.

To assess the validity of the approximations used in the analytical model across different magnetostrictive layer thicknesses, three additional simulations were conducted for magnetostrictive layer thicknesses of between 20 nm and 40 nm. These particular values were selected as they are representative of spin-wave waveguide dimensions, and the transducer efficiency begins to exhibit saturation at these thicknesses. The simulation was performed solely for CoFeB (with a $Q$-factor of 100) due to the computationally intensive nature of the task: for each magnetostrictive layer thickness, the externally applied magnetic field must be tuned to overlap FMR with the mechanical resonance, which adds significant computational overhead. The results, presented in Fig.~\ref{eff_ifo_t}(a), demonstrate good agreement between the analytical and FEM models, validating the accuracy of the analytical approximations for the given range of magnetostrictive layer thicknesses.

\subsubsection{Influence of the magnetoelastic coupling constant}

\begin{figure}[p] 
	\centering
	\includegraphics[scale=0.7]{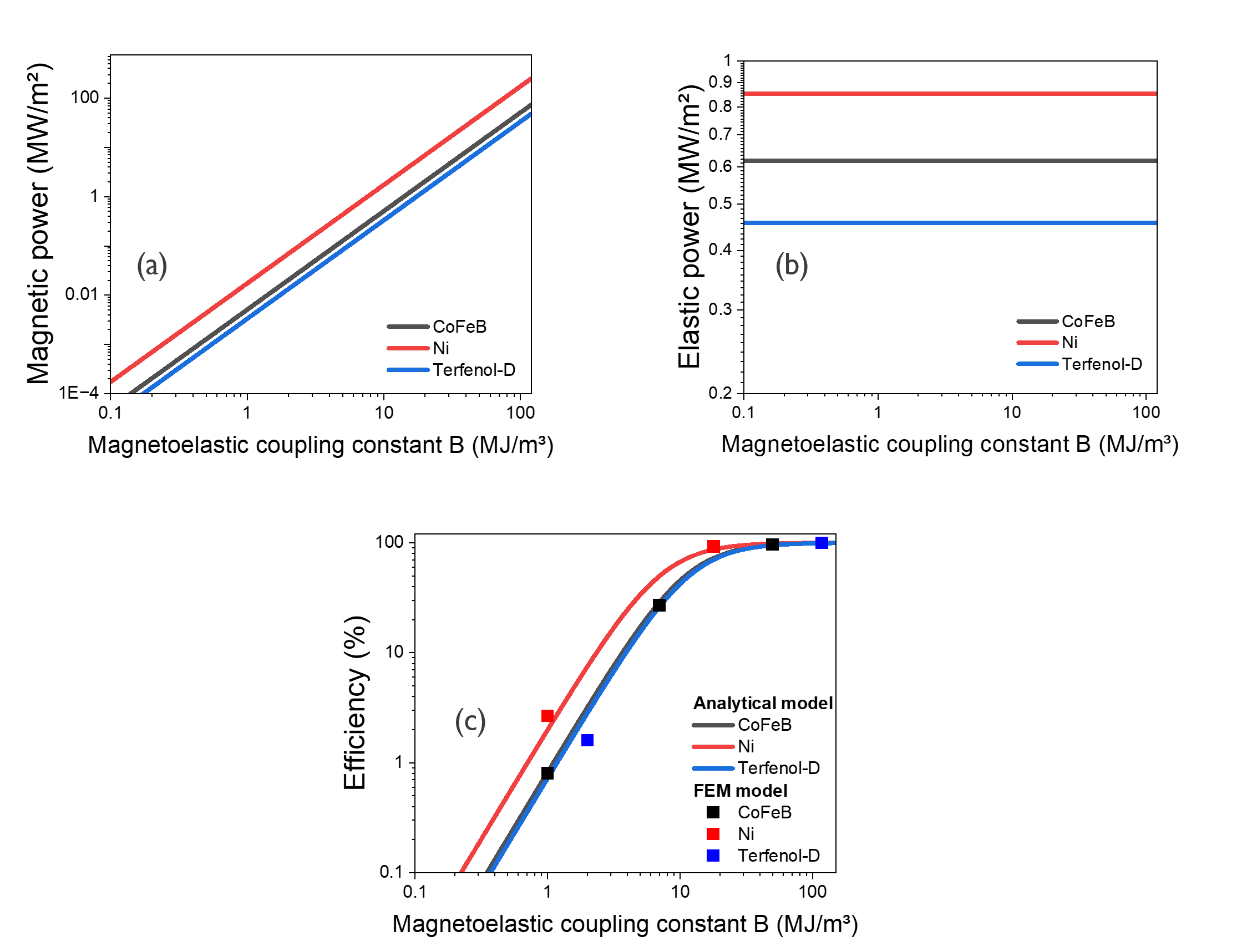}
	\caption{(a) Magnetic power absorption $P_{m}$, (b) elastic power loss $P_{e}$, and (c) magnetic transduction efficiency $\eta$ as functions of the magnetoelastic coupling constant $B$, evaluated for material parameter sets corresponding to CoFeB, Ni, and Terfenol-D. All results are computed for a ScAlN resonator with mechanical quality factor $Q = 1000$. Analytical model predictions are shown as solid lines, while corresponding FEM results are indicated by squares. The comparison highlights good agreement between the two approaches.}
	\label{eff_ifo_B}
\end{figure}

Next, we examine the influence of the magnetoelastic coupling constant $B$ on the power absorption characteristics and transduction efficiency of the magnetoelectric resonator. In this analysis, as well as in the subsequent section, the thickness of the magnetostrictive layer is fixed at 20 nm. To isolate the effect of $B$, all other material and geometric parameters are held constant. Furthermore, the mechanical quality factor is $Q = 1000$ in all cases. Figure~\ref{eff_ifo_B} presents the calculated magnetic and elastic power dissipation, along with the corresponding transduction efficiency, as functions of $B$.

The magnetic power absorption exhibits a quadratic dependence on the magnetoelastic coupling constant, as shown in Fig.~\ref{eff_ifo_B}(a). This behavior is consistently observed across all three material systems studied. Notably, Ni demonstrates the highest magnetic power absorption at all given $B$, which can be attributed to its relatively low saturation magnetization, enhancing the magnetoelastic effective field (\textit{cf.} Eq.~\eqref{Eq:Hmel}). In contrast, the elastic power loss is rather independent of $B$, as expected [see Fig.~\ref{eff_ifo_B}(b)]. Among the materials considered, Ni again shows the highest elastic losses due to its larger elastic stiffness. As illustrated in Fig.~\ref{eff_ifo_B}(c), the resulting transduction efficiency increases monotonically with $B$, approaching saturation near 100\% for $ B > 10$ MJ/m$^3$. This trend provides insight into the high efficiencies observed in Ni- and Terfenol-D–based resonators, as well as the comparatively lower efficiency in CoFeB systems. 

To validate the analytical predictions, FEM simulations were conducted to compute the transduction efficiency as a function of $B$ for all three material systems. The results in Fig.~\ref{eff_ifo_B}(c) demonstrate good agreement between the analytical and FEM models, particularly in the case of CoFeB, thereby reinforcing the reliability of the analytical approach within the investigated parameter space.

\subsubsection{Influence of the magnetic Gilbert damping}

\begin{figure}[p] 
	\centering
	\includegraphics[scale=0.7]{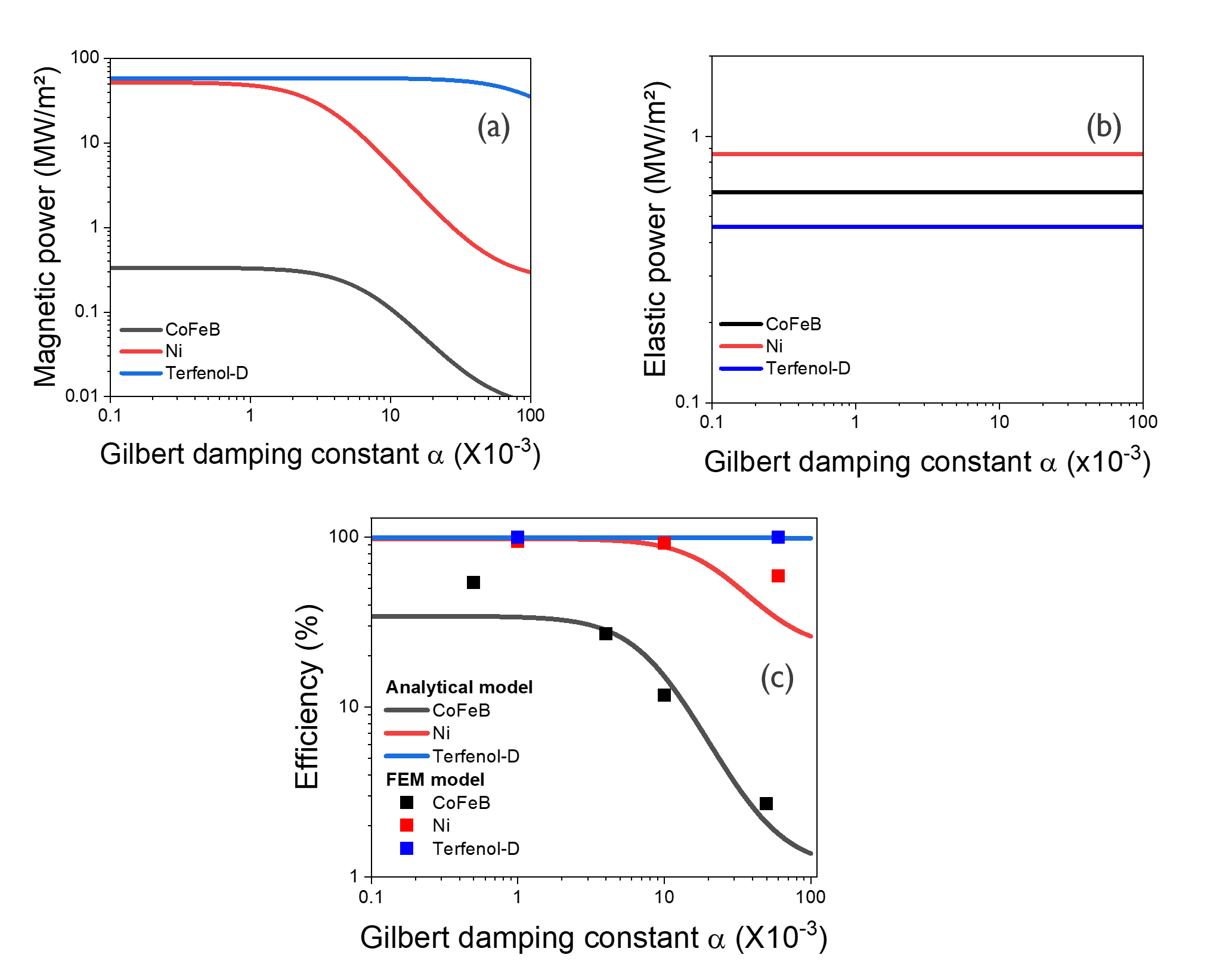}
	\caption{(a) Magnetic power absorption $P_{m}$, (b) elastic power loss $P_{e}$, and (c) magnetic transduction efficiency $\eta$ as functions of the Gilbert damping constant $\alpha$, evaluated for material parameters corresponding to CoFeB, Ni, and Terfenol-D. All results are computed for a ScAlN resonator with mechanical quality factor $Q = 1000$. Results from the analytical model are shown as solid lines, while corresponding FEM results are represented by squares. Again, good overall agreement between the two approaches is observed.}
	\label{eff_ifo_a}
\end{figure}

We now investigate the influence of the Gilbert damping constant $\alpha$ on the magnetoelectric resonator's performance. As in the preceding analysis, we evaluate the magnetic and elastic power dissipation, as well as the resulting transduction efficiency, while varying $\alpha$ and holding all other material and geometric parameters constant. The results of this parametric study are presented in Fig.~\ref{eff_ifo_a}.

Figure~\ref{eff_ifo_a}(a) reveals that, at a mechanical quality factor of $Q = 1000$ and at low values for $\alpha$, the magnetic power loss remains constant. At $\alpha > 10^{-3}$, the magnetic power absorption and thus the transduction efficiency starts to decrease. This decrease is first seen for Ni, due to its low saturation magnetization. For all damping values, Terfenol-D shows the highest magnetic power absorption due to its large magnetoelastic coupling constant, while CoFeB consistently displays the lowest magnetic losses for all $\alpha$, due to its comparatively small magnetoelastic coupling and high saturation magnetization.

As shown in Fig.~\ref{eff_ifo_a}(b), the elastic power dissipation remains unaffected by changes in the magnetic damping constant. This is consistent with the assumption of negligible backaction from the magnetic subsystem to the elastic domain in the considered model. Among the materials, Ni again exhibits the highest elastic losses, driven by its relatively high elastic stiffness.

The calculated transduction efficiencies, shown in Fig.~\ref{eff_ifo_a}(c), increase with decreasing $\alpha$, reflecting the enhanced magnetic energy absorption at lower damping. Terfenol-D achieves the highest efficiencies across the entire $\alpha$ range, owing to the combination of strong magnetoelastic coupling and relatively modest elastic losses. In contrast, CoFeB remains the least efficient, primarily due to its weaker magnetoelastic interaction. Finally, the FEM simulation results show good agreement with the analytical predictions across all three material systems and for the entire range of $\alpha$, further validating the accuracy and robustness of the analytical model in capturing damping-dependent behavior.

\subsubsection{Influence of the mechanical \textit{Q}-factor of the resonator}

\begin{figure}[p] 
	\centering
	\includegraphics[scale=0.7]{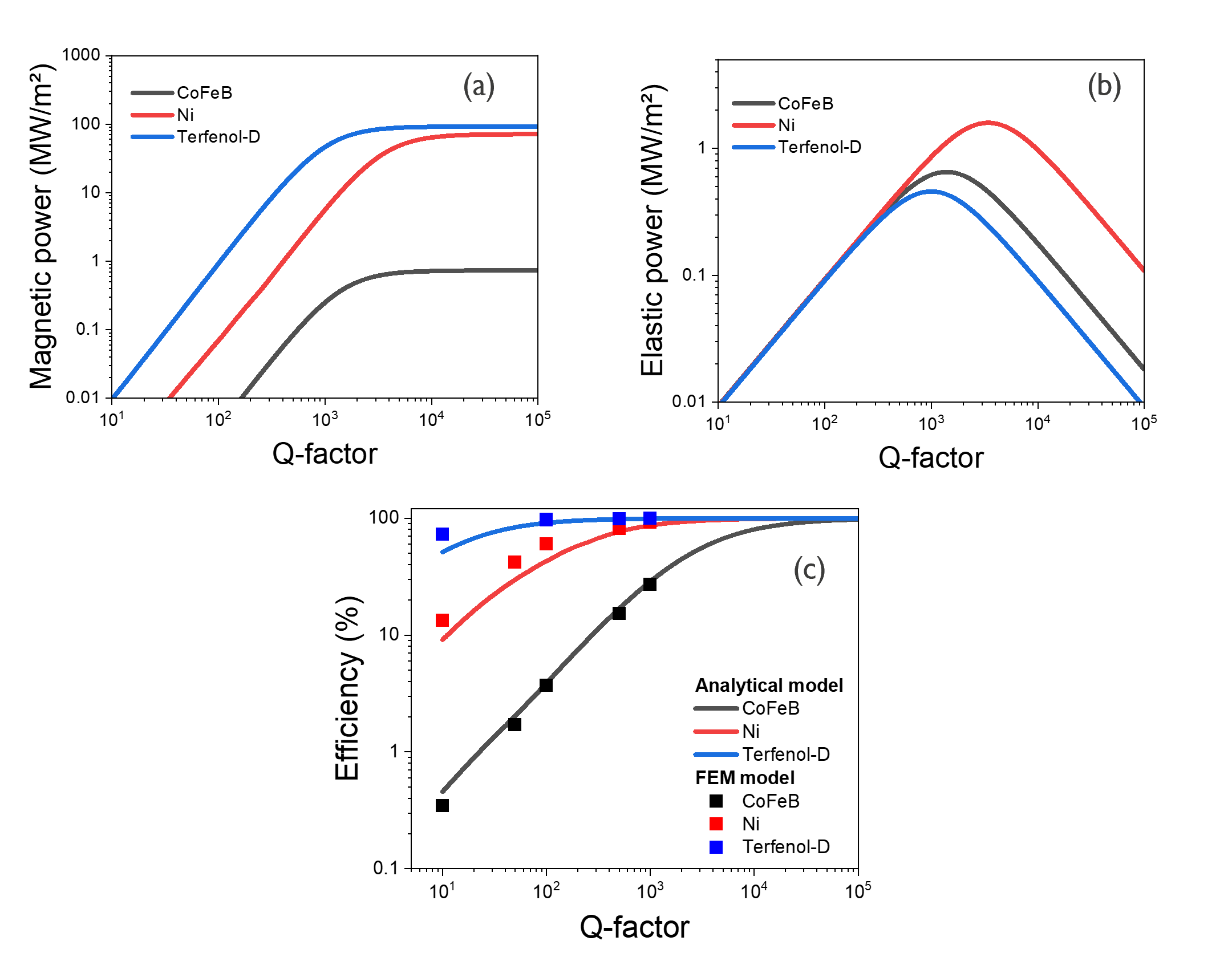}
	\caption{(a) Magnetic power absorption $P_{m}$, (b) elastic power loss $P_{e}$, and (c) magnetic transduction efficiency $\eta$ as functions of the mechanical $Q$-factor for material parameters corresponding to CoFeB, Ni, and Terfenol-D. Results from the analytical model are shown as solid lines, while corresponding FEM results are represented by squares.}
	\label{eff_ifo_Q}
\end{figure}

Finally, we analyze the influence of the mechanical quality factor $Q$ on magnetic and elastic power dissipation, as well as on the overall transducer efficiency. The results of this parametric study are presented in Fig.~\ref{eff_ifo_Q}.

Figure~\ref{eff_ifo_Q}(a) shows that magnetic power absorption increases sharply with increasing $Q$, eventually saturating for $Q \gtrsim 1000$. In contrast, the elastic power dissipation in Fig. \ref{eff_ifo_Q}(b) initially increases with $Q$, reaches a maximum, and subsequently decreases. This non-monotonic behavior arises due to the $Q$-dependent strain amplitude within the piezoelectric layer. At low $Q$, elastic losses dominate due to the relatively high imaginary component of the piezoelectric stiffness. As $Q$ increases (\textit{i.e.}, mechanical losses are reduced), the imaginary component of the effective stiffness diminishes, enhancing the strain amplitude. This enhanced strain leads to a quadratic increase in magnetic power dissipation, consistent with Eq.~\eqref{total_magnetic_power}, since $P_{m} \propto S_0^{m}S_{k}^{m}$.

Elastic power loss also scales quadratically with strain, but is modulated by the imaginary part of the piezoelectric stiffness, which itself scales inversely with $Q$, according to $P_{el} \propto c_{i}^{p}S^{2} \propto S^{2}/Q$ [see Eq.~\eqref{eq_elastic_power}]. Consequently, while the elastic loss initially increases due to the increasing strain, it eventually decreases because the stiffness damping term vanishes more rapidly with increasing $Q$. At sufficiently high $Q$, magnetic losses dominate the overall dissipation, determining the strain amplitude, which then becomes independent of $Q$. As a result, the magnetic power absorption saturates, while elastic losses continue to decline. This dynamic explains the observed peak in elastic power as a function of $Q$. Note that this qualitative behavior is largely independent of the specific magnetostrictive material parameters and is expected to be generally applicable.

The corresponding transduction efficiency $\eta$ is plotted in Fig.~\ref{eff_ifo_Q}(c). Because the magnetic power absorption increases rapidly and eventually dominates, while elastic losses plateau or decrease, the efficiency increases monotonically with $Q$ and approaches unity at large $Q$. Notably, high efficiencies exceeding 50\% are already achieved for Ni and Terfenol-D at moderate $Q$-factors near 100. Figure~\ref{eff_ifo_Q}(c) also highlights the good agreement between analytical predictions and FEM simulations across all considered materials and over the full numerically accessible $Q$-range. Due to the computational expense associated with resolving sharp resonances at high $Q$, FEM simulations were restricted to $Q \le 1000$.

\section{Conclusion}

In conclusion, we have developed a comprehensive analytical model to describe power transduction in a magnetoelectric resonator comprising a piezoelectric--magnetostrictive bilayer. By deriving and solving the coupled differential equations that govern the dynamics of the elastic and magnetic subsystems under free boundary conditions, we obtained a general solution for the strain and magnetization dynamics. These solutions were then used to derive closed-form expressions for magnetic and elastic power dissipation as functions of geometrical and material parameters. From these expressions, the magnetic transduction efficiency could be defined and quantified, revealing its dependence on the strain amplitude within the device.

To validate the analytical framework and assess the impact of model assumptions, a FEM model of the bilayer resonator was implemented. By systematically comparing the analytically and numerically computed efficiencies across a broad range of magnetostrictive layer thicknesses, magnetoelastic coupling coefficients, Gilbert damping parameters, and mechanical quality factors $Q$, good agreement between both models was observed. This confirms the negligible impact of magnetic backaction on the elastic subsystem in the studied configurations, a consequence of the dominance of exchange interactions in the thin magnetic films, resulting in spatially uniform magnetization dynamics. Moreover, the FEM analysis enabled time-domain investigations of resonator dynamics, revealing extended transient regimes at high $Q$-factors and delineating the linear operational limits under applied electric fields.

Using both models, a detailed case study was conducted on bilayer structures combining piezoelectric ScAlN with representative magnetostrictive materials (CoFeB, Ni, and Terfenol-D) to elucidate the influence of material and geometrical parameters on power dissipation and transduction efficiency. A key insight from this study is the critical role of the magnetoelastic coupling coefficient as well as the saturation magnetization in maximizing transduction efficiency. Terfenol-D and Ni were shown to achieve hight transduction efficiencies approaching 100\% at moderate $Q$-factors ($Q \ge 100$), while efficiencies above 10\% are attainable even at lower $Q$-factors around $Q \approx 10$ for 20 nm magnetic layer thicknesses. In contrast, CoFeB, despite its favorable damping properties, exhibits reduced efficiency due to its combination of low magnetoelastic coupling coefficient and high saturation magnetization. These high values, particularly for Ni and Terfenol-D, indicate that these devices can still be efficienct in regimes where fabrication-related imperfections or environmental factors limit the mechanical $Q$-factors to values as low as 10–100. This makes the devices viable for practical integration into real-world nanoscale systems.

These findings demonstrate the strong potential of magnetoelectric transducers to achieve high-efficiency magnetic excitation, such as FMR and spin waves, which are central to a wide range of emerging spintronic and magnonic technologies \cite{15_KhitunAlexander2009Mswa,39_KhitunAlexander2007Fsol,5_KhitunAlexander2011Nmlc}. An important feature of the proposed resonator concept is the area-independence of the efficiency (assuming idealized boundary conditions and negligible edge or corner effects). This characteristic distinguishes magnetoelectric resonators from conventional inductive transducers, which typically suffer from decreased performance as device dimensions are scaled down. Consequently, magnetoelectric transducers are particularly well suited for use in highly miniaturized platforms, including for on-chip microwave signal processing, localized spin-wave generation, and hybrid quantum devices that require efficient strain-to-magnetization conversion at the nanoscale.

Furthermore, the analytical framework developed in this work provides a computationally efficient predictive tool for evaluating and optimizing device performance over a wide parameter space. It enables rapid screening of candidate material combinations and geometries. As demonstrated in the case study, the model can be readily applied to compare the performance of multiple magnetostrictive materials. By contrast, the FEM enables direct verification of the assumptions and approximations underlying the analytical formulation. Furthermore, the FEM approach can serve as a foundation for future extensions, \textit{e.g.}, to explore nonlinear effects, or the impact of structurally complex or multilayered geometries. It also facilitates the inclusion of magnetic anisotropy, spatially varying material properties, and interfacial phenomena, factors that are difficult to incorporate into closed-form analytical models.

\section{Acknowledgement}

This work was supported by imec’s Industrial Affiliate Program on Beyond-CMOS Logic. Additional funding was provided by the European Union’s Horizon 2020 Research and Innovation Program under the FET-OPEN project CHIRON (Grant Agreement No. 801055). E.V.M. and F.V. gratefully acknowledge financial support from the Research Foundation -- Flanders (FWO) through grants No. 1SH4Q24N and 1S05719N, respectively. The authors also wish to thank Roman Verba for his perspective and insightful discussions.
\clearpage
\appendix

\section{Magnetoelastic stress and body force\label{App:A}}

The magnetoelastic energy for a material exhibiting cubic (or higher) crystal symmetry can be expressed as \cite{58_magnetizationoscillation,59_alma99554810101488} 
\begin{equation}
    E_{mel} = \int_V \left[ \frac{B_{1}}{M_{S}^{2}}\sum_{i} M_{i}^{2}S_{ii} + B_{2} \sum_{i\neq j} M_{i}M_{j}S_{ij}\\
    \right]\,dV .
\end{equation}
\noindent Here, $S_{ij}$ represents the components of the strain tensor, while $B_{1,2}$ are the magnetoelastic coupling constants. Consequently, the magnetoelastic stress tensor is given by
\begin{equation}
    \sigma_{mel,ij} = \frac{\partial E_\mathrm{mel}}{\partial S_{ij}}
\end{equation}
\noindent Considering that only the shear strain $S_{xz} = S_{zx}$ is nonzero---consistent with the resonator configuration used in this study---results in nonzero shear stresses $\sigma_{zx}=\sigma_{xz}$. The expression for this stress component is then given by
\begin{equation}
    \sigma_{\mathrm{mel},xz} = \frac{B_{2}}{M_{S}^{2}}M_{x}M_{z}
\end{equation}
\noindent The body force is obtained by taking the derivative of the stress tensor, which results in  \cite{60_VandervekenFrederic2020Cmwi,61_alma9910173270101488,62_alma9911509040101488}
\begin{equation}
    f_{\mathrm{mel},i} = \frac{\partial}{\partial x_{j}} \frac{\delta E_{tot}}{\delta S_{ij}}
\end{equation}

\section{Magnetization dynamics as a function of strain\label{App:B}}
In this appendix, the amplitude of the magnetization dynamics is derived from the strain amplitude within the magnetostrictive layer. The interaction between the strain and the magnetization are described by the magnetoelastic field, which is given by
\begin{align}
    H_\mathrm{mel} &= -\frac{1}{\mu_{0}}\frac{\delta E_{mel}}{\delta M} \\
    &=-\frac{2}{\mu_{0}M_{S}^{2}}\begin{bmatrix}
        B_{1}S_{xx}M_{x} + B_{2}\left(S_{xy}M_{y} + S_{zx}M_{z}\right)\\
        B_{1}S_{yy}M_{y} + B_{2}\left(S_{xy}M_{x} + S_{yz}M_{z}\right)\\
        B_{1}S_{zz}M_{z} + B_{2}\left(S_{zx}M_{x} + S_{yz}M_{y}\right)\\
    \end{bmatrix}. \label{Eq:Hmel}
\end{align}
\noindent For the given geometry and configuration, only the shear strain component $S_{xz}=S_{zx}$ is nonzero, resulting in
\begin{equation}
    =-\frac{2B_{2}S_{zx}^{2}}{\mu_{0}M_{S}^{2}}\begin{bmatrix}
        M_{z}\\
        0\\
        M_{x}\\
    \end{bmatrix}.
    \label{eq_resultingHmel}
\end{equation}
\noindent For clarity and improved readability, the subscripts are omitted in the following, such that $B_{2}\equiv B$ and $S_{zx}^{m}\equiv S^{m}$.

The effective magnetic field is consists of several contributions, including the exchange field, the demagnetization field, and the externally applied field $H_{0}$. As discussed in the main text in detail, the magnetic excitation can be considered uniform across the thickness of the magnetostrictive film. This approximation justifies neglecting the exchange interaction. The demagnetization field can then be approximated by
\begin{equation}
    H_{d} = \begin{bmatrix}
        0\\
        0\\
        -M_{z}\\
    \end{bmatrix}.
\end{equation}
\noindent This results in the total effective field
\begin{equation}
    H_\mathrm{eff} = \begin{bmatrix}
        H_{0} - \frac{2BS^{m}}{\mu_{0}M_{S}^{2}}M_{z}\\
        0\\
        -M_{z} - \frac{2BS^{m}}{\mu_{0}M_{S}^{2}}M_{x}\\
    \end{bmatrix}.
\end{equation}
\noindent Substituting the effective magnetic field into the Landau–Lifshitz–Gilbert (LLG) equation and subsequently linearizing the resulting differential equation yields
\begin{equation}
    i\omega \begin{bmatrix}
        M_{y}\\
        M_{z}\\
    \end{bmatrix} = \begin{bmatrix}
        -(\omega_{0}+\omega_{M})M_{z} - 2BS^{m}\gamma\\
        \omega_{0}M_{y}\\
    \end{bmatrix} + i\omega \alpha \begin{bmatrix}
        -M_{z}\\
        M_{y}\\
    \end{bmatrix}.
\end{equation}
\noindent These equations can be rewritten to find relations between the magnetization components and the strain
\begin{align}
    M_{y} &= -i \frac{2B\gamma \omega}{\omega_{y}\omega_{z}-\omega^{2}}S^{m} \\
    &= -i \frac{2B\gamma \omega}{\omega_{r}^{2}-\omega^{2}+i\alpha \omega (2\omega_{0}+\omega_{M})}S^{m}, \\
    M_{z} &= -i \frac{2B\gamma \omega_{y}}{\omega_{y}\omega_{z}-\omega^{2}}S^{m} \\
    &= -i \frac{2B\gamma \omega_{y}}{\omega_{r}^{2}-\omega^{2}+i\alpha \omega (2\omega_{0}+\omega_{M})}S^{m},
\end{align}
\noindent with $\omega_{y}=\omega_{0}+i\omega \alpha$ and $\omega_{z}=\omega_{0} + \omega_{M} + i\omega \alpha$. Assuming that $\alpha \ll 1$, we can write
\begin{equation}
    \omega_{y}\omega_{z} \approx \omega_{0}(\omega_{0}+\omega_{M}) + i\omega\alpha (2\omega_{0}+\omega_{M}) = \omega_{r}^{2} + i\omega \alpha (2\omega_{0}+\omega_{M}),
\end{equation}
\noindent with $\omega_{r}^{2}=\omega_{0}(\omega_{0} + \omega_{M})$ the magnetic resonance frequency. 

\section{Strain-induced spatially-uniform magnetization dynamics in the magnetostrictive layer\label{App:C}}
\noindent In this appendix, we evaluate the effective strain within the magnetostrictive layer that governs the excitation of magnetization dynamics. Since $k_{m}t \ll 1$, the long-wavelength approximation is valid, allowing us to treat the magnetization as spatially uniform across the magnetostrictive layer. Consequently, the effective strain can be approximated in the limit $k \rightarrow 0$. The strain expression in the magnetostrictive layer, as defined in Eq.~(\ref{eq:Strain_M}), can be transformed into the wavevector domain via the Fourier integral
\begin{align}
	S^{m}(k) &= {eE}\frac{k^{m}\sin^{2}\!\left(\frac{1}{2}k_{t}^{p}d\right)}{c^{m}k^{m}\cos (k_{t}^{p}d )\sin\!\left(k^{m}t\right) + c_{t}^{p}k_{t}^{p}\sin\!\left(dk_{t}^{p}\right)\cos (k^{m}t )} \int_{-t}^{0}\sin\!\left(k^{m}(t+z)\right)e^{-ikz}dz,
	\label{eqC1:Strain_M}\\
&=\frac{eE}{t}\frac{k^{m}\sin^{2}\!\left(\frac{1}{2}k_{t}^{p}d\right)(k^{m}e^{ikt}  - ik\sin (k^{m}t) - k^{m}\cos (k^{m}t)))}{(c^{m}k^{m}\cos (k_{t}^{p}d )\sin\!\left(k^{m}t\right) + c_{t}^{p}k_{t}^{p}\sin\!\left(dk_{t}^{p}\right)\cos (k^{m}t ))((k^{m})^{2}-k^{2})} .
	\label{eqC2:Strain_M}
\end{align}
\noindent Note that $S^{m}(k)$ exhibits a sinc-like profile with a maximum at $k = 0$. In the limit of $k = 0$, Eq. \eqref{eqC2:Strain_M} becomes
\begin{equation}
	S^{m}_0 \equiv S^{m}(k=0) = \frac{eE}{ t}\frac{k^{m}\sin^{2}\!\left(\frac{1}{2}k_{t}^{p}d\right)(k^{m} - k^{m}\cos(k^{m}t)))}{(c^{m}k^{m}\cos (k_{t}^{p}d )\sin\!\left(k^{m}t\right) + c_{t}^{p}k_{t}^{p}\sin (dk_{t}^{p} )\cos (k^{m}t ))(k^{m})^{2}} .
	\label{eqC3:Strain_M}
\end{equation}

\clearpage

\bibliographystyle{apsrev}
\bibliography{biblio}

\clearpage

\clearpage

\begin{table}[p]
	\begin{tabular}{l!{\hspace{0.5cm}}c!{\hspace{0.35cm}}c!{\hspace{0.35cm}}c!{\hspace{0.35cm}}c!{\hspace{0.35cm}}c!{\hspace{0.35cm}}c!{\hspace{0.35cm}}}
		\hline		\hline
		Material & $\rho$ ($g/cm^{3}$) & $c_{0}$ (GPa)  & $e$ (C/m$^{2}$) & $B$ (MJ/m$^{3}$) & $M_{S}$  (kA/m) & $\alpha$  ($\times 10^{-3}$) \\
		\hline
		ScAlN & 3.5 & 100 & 0.3 & -- & -- & -- \\
		CoFeB & 8.0 & 70 & -- & 7 & 1000 & 4\\
		Ni & 8.9 & 74 & -- & 18 & 370 & 10 \\
		Terfenol-D & 9.3 & 38 & -- & 118 & 700 & 60 \\
		\hline
		\hline
	\end{tabular}
	\caption{Material parameters for ScAlN \cite{51_CaroMiguelA2015Pcas}, CoFeB \cite{53_PengRenCi2016F1ms}, Ni \cite{54_DreherL2012SAWF, magnetostriction_Ni_AnoopBabyK.B.2022Msot} and Terfenol-D (Tb$_{0.3}$Dy$_{0.7}$Fe$_{2}$) \cite{55_ColussiMarco2016SEDB,56_gopman,57_WangQianchang2017S1si}: mass density $\rho$, bulk stiffness constant $c_{0}$, piezoelectric constant $e$, magnetoelastic coupling constant $B$, saturation magnetization $M_{S}$, and Gilbert damping constant $alpha$. These parameters are consistently employed in both the analytical and FEM models to ensure direct comparability.}
	\label{table:1}
\end{table}

\begin{table}[p]
	\begin{tabular}{l!{\hspace{0.5cm}}c!{\hspace{0.35cm}}c!{\hspace{0.35cm}}}
		\hline	\hline
		Material & External field $\mu_0H$ (mT) & Resonance frequency $\nu_\mathrm{res}$ (GHz) \\ 
		\hline
		CoFeB & 115 & 11.0 \\
		Ni & 219 & 10.8 \\
		Terfenol-D & 179 & 10.0 \\
		\hline
		\hline
	\end{tabular}
	\caption{Externally applied magnetic field $\mu_{0}H$ and corresponding ferromagnetic-–mechanical resonance frequency $\nu_{\text{res}}$ for ScAlN-based FBARs incorporating magnetostrictive CoFeB, Ni, and Terfenol-D layers. The piezoelectric and magnetostrictive layer thicknesses are fixed at 200 nm and 20 nm, respectively. These parameter sets are consistently used in both the analytical and FEM models.\label{table:2}}
\end{table}

\end{document}